%% file: main.tex
\documentclass[sigconf,pbalance]{acmart}

\copyrightyear{2026}
\acmYear{2026}
\setcopyright{cc}
\setcctype{by}
\acmConference[CHI '26]{Proceedings of the 2026 CHI Conference on Human Factors in Computing Systems}{April 13--17, 2026}{Barcelona, Spain}
\acmBooktitle{Proceedings of the 2026 CHI Conference on Human Factors in Computing Systems (CHI '26), April 13--17, 2026, Barcelona, Spain}
\acmPrice{}
\acmDOI{10.1145/3772318.3791165}
\acmISBN{979-8-4007-2278-3/2026/04}

\acmSubmissionID{9946}

\input{preamble}

\begin{document}

\title{Improving Low-Vision Chart Accessibility via On-Cursor Visual Context}

\include{authors}

\begin{abstract}
  \input{content/00-abstract}
\end{abstract}



\begin{teaserfigure}
  \includegraphics[width=\textwidth]{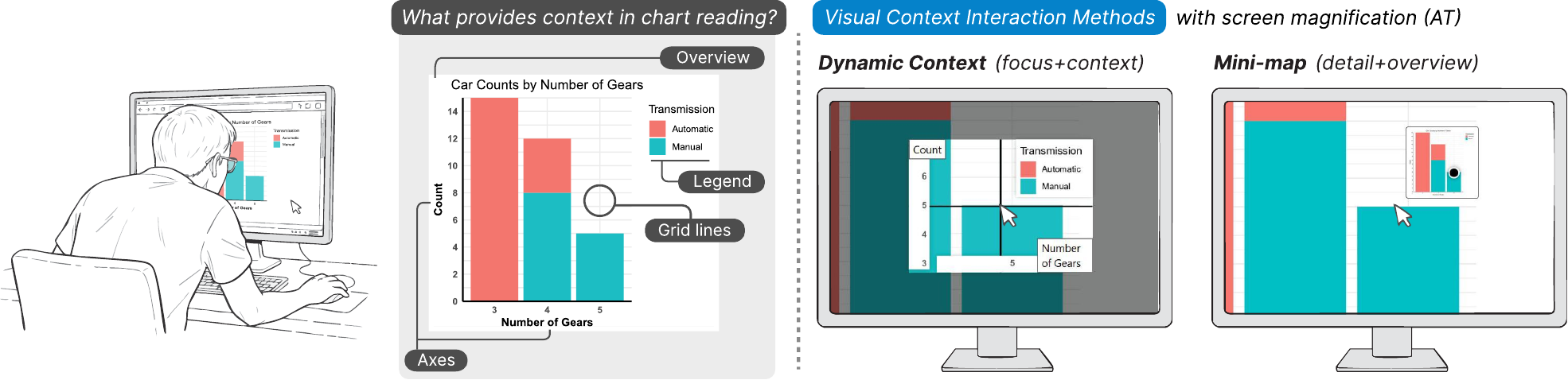}
  \caption{Our proposed interaction methods provide access to visual context elements for low-vision individuals (LVI). \dc is inspired by focus+context approaches, and \mm is inspired by overview+detail approaches. Both interaction methods are designed to integrate with existing assistive tools (AT), e.g., screen magnification. }
  \Description{An illustration of a person with low vision leaning towards the monitor, trying to read a chart (left). The chart is shown in full with the caption "What provides context in chart reading?". Four highlighted chart elements are: axes, legend, grid lines, and the overview. Both interaction methods are shown side-by-side. Dynamic Context shows the axes, legend, and one horizontal and vertical grid lines which connect at the center of the display. Mini-map shows a small floating copy of the chart with an indication of the current position relative to the chart.}
  \label{fig:system-teaser}
\end{teaserfigure}


\maketitle

\input{content/01-introduction}

\input{content/02-background}

\input{content/03-formative}

\input{content/04-implementation}

\input{content/05-user-study}

\input{content/06-discussion}

\input{content/07-conclusion}

\begin{acks}
    \input{content/11-acknowledgment}
\end{acks}

\bibliographystyle{ACM-Reference-Format}
\bibliography{ref.global, ref.main}

\appendix

\include{content/99-appendix}

\end{document}

%% file: preamble.tex
\usepackage{pifont}
\newcommand{\cmark}{\ding{51}}
\newcommand{\xmark}{\ding{55}}
\usepackage{xcolor}
\usepackage{multirow}
\usepackage{makecell}
\usepackage{subcaption}
\usepackage{xspace}
\usepackage{enumitem}
\newcolumntype{L}[1]{>{\raggedleft\arraybackslash}p{#1}}
\newcolumntype{R}[1]{>{\raggedright\arraybackslash}p{#1}}
\newcommand{\q}[1]{\textit{``#1''}}
\newcommand{\A}{Appendix~}
\newcommand{\eg}{e.g.,\xspace}
\newcommand{\bl}{baseline\xspace}
\newcommand{\mm}{Mini-map\xspace}
\newcommand{\dc}{Dynamic Context\xspace}

%% file: authors.tex

\settopmatter{authorsperrow=4}

\author{Yotam Sechayk}
\affiliation{%
  \institution{The University of Tokyo}
  \city{Tokyo}
  \country{Japan}
}
\orcid{0009-0002-5286-0080}
\email{sechayk-yotam@g.ecc.u-tokyo.ac.jp}

\author{Hennes Rave}
\affiliation{%
  \institution{University of Münster}
  \city{Münster}
  \country{Germany}
  }
\orcid{0000-0001-8662-3767}
\email{hennes.rave@uni-muenster.de}

\author{Max R\"adler}
\affiliation{%
  \institution{Ulm University}
  \city{Ulm}
  \country{Germany}
  }
\orcid{0000-0002-5413-2637}
\email{max.raedler@uni-ulm.de}

\author{Mark Colley}
\affiliation{%
  \institution{University College London}
  \city{London}
  \country{United Kingdom}
  }
\orcid{0000-0001-5207-5029}
\email{m.colley@ucl.ac.uk}

\author{Zhongyi Zhou}
\affiliation{%
  \institution{Google}
  \city{Tokyo}
  \country{Japan}
  }
\orcid{0000-0001-8908-1086}
\email{zhongyizhou@google.com}

\author{Ariel Shamir}
\affiliation{%
  \institution{Reichman University}
  \city{Herzliya}
  \country{Israel}
  }
\orcid{0000-0001-7082-7845}
\email{arik@runi.ac.il}

\author{Takeo Igarashi}
\affiliation{%
  \institution{The University of Tokyo}
  \city{Tokyo}
  \country{Japan}
}
\orcid{0000-0002-5495-6441}
\email{takeo@acm.org}

\renewcommand{\shortauthors}{Sechayk et al.}

%% file: content/00-abstract.tex
Despite widespread use, charts remain largely inaccessible for Low-Vision Individuals (LVI). Reading charts requires viewing data points within a global context, which is difficult for LVI who may rely on magnification or experience a partial field of vision. We aim to improve exploration by providing visual access to critical context. To inform this, we conducted a formative study with five LVI. We identified four fundamental contextual elements common across chart types: axes, legend, grid lines, and the overview. We propose two pointer-based interaction methods to provide this context: Dynamic Context, a novel focus+context interaction, and Mini-map, which adapts overview+detail principles for LVI. In a study with N=22 LVI, we compared both methods and evaluated their integration to current tools.  Our results show that Dynamic Context had significant positive impact on access, usability, and effort reduction; however, worsened visual load. Mini-map strengthened spatial understanding, but was less preferred for this task. We offer design insights to guide the development of future systems that support LVI with visual context while balancing visual load.

%% file: content/01-introduction.tex
\section{Introduction}
\label{sec:introduction}

Data charts are ubiquitous in communicating information~\cite{in2017StatisticalData,alcaraz-martinez2024EnhancingStatistical}.
However, charts contain rich visual information and require the user to access fine-grained details like data points and their broader context~\cite{kim2021AccessibleVisualization,marriott2021InclusiveData}. 
For instance, making sense of stacked bar charts hinges on connecting bar edges to axis values, relating colors to the legend, and comparing two or more bars.
Doing this is especially difficult for people with vision impairments~\cite{wang2024HowLowvision,prakash2024UnderstandingLow,alcarazmartinez2022MethodologyHeuristic,zong2022RichScreen}, despite their growing prevalence. 
Global data~\cite{bourne2021trends} estimate that 553 million people lived with low vision in 2020, a figure predicted to increase substantially as moderate and severe impairment alone is projected to reach 474 million people by 2050.
For low-vision individuals (LVI), accessing charts remains a challenge~\cite{wang2024HowLowvision, prakash2024UnderstandingLow, alcaraz-martinez2024EnhancingStatistical}. 
Limited access diminishes information equity, specifically by restricting rich visual context and data insights derived from visualizations~\cite{marriott2021InclusiveData}---such as during health crises~\cite{fan2023AccessibilityData}.

Current strides in making charts accessible for people with visual impairments are centered around non-visual access, such as via screen readers~\citep[\eg][]{seo2024MAIDRMaking,moured2024Chart4BlindIntelligent,blanco2022OlliExtensible,nazemi2013MethodProvide,choi2019VisualizingNonVisual,elavsky2024DataNavigator,ferres2013EvaluatingTool}, tactile/haptic~\citep[\eg][]{gotzelmann2016LucentMaps3D,milallos2024ExploringEffectiveness,moured2024Chart4BlindIntelligent,singh2023GraphoBringing}, or sonification techniques~\citep[\eg][]{zhao2024TADAMaking,chundury2024TactualPlotSpatializing}. 
However, while many works group blind with LVI~\cite{mack2021WhatWe,marriott2021InclusiveData}, the latter prefer using their residual vision to access visual media~\cite{szpiro2016HowPeople}. Therefore, chart accessibility should go beyond non-visual access~\cite{wimer2024VisionImpairments,marriott2021InclusiveData} and support LVI in interpreting data independently~\cite{keilers2023DataVisualization}.
Assistive tools (AT) that support visual access for LVI are largely limited to magnification, which provides access to details. However, it creates a trade-off between detail and context.
With full-screen magnification, content outside the zoomed area disappears~\cite{szpiro2016HowPeople,islam2023SpaceXMag}. 
Conversely, lens magnification shows the zoomed section within the full screen; however, the lens itself occludes nearby content.
Therefore, it is common for LVI to constantly pan to different regions of a chart or adjust zoom levels, relying on memory and constructing mental models to access chart information~\cite{theofanos2005HelpingLowvision}.
This becomes increasingly demanding for complex charts, such as charts with numerous categories or data series~\cite{alcaraz-martinez2024EnhancingStatistical}. 
Our aim was to effectively deliver context within existing AT usage to support chart reading and exploration, and preserve the visual agency of LVI.

To explore how LVI exercise visual agency and use context when reading charts, we conducted a formative study with five participants to understand their visual access strategies, AT usage, and prioritized contextual elements.
For a broad coverage of charts, we used a chart comprehension test designed with 12 common chart types~\cite{lee2017VLATDevelopment,pandey2023MiniVLATShort}.
Our findings confirm that LVI prefer visual access using AT~\cite{szpiro2016HowPeople}, specifically magnification and color filters. Current strategies require LVI to dedicate substantial effort to panning and zooming across charts to assimilate the structure and access the underlying data~\cite{wang2024HowLowvision,prakash2024UnderstandingLow}. 
We surface four contextual elements---most common across chart types---that LVI use for accessing this data: legends, axes, grid lines, and the overall view. 
However, locating axes and legends often requires guesswork, grid lines frequently lack the contrast needed for accurate tracking, and accessing the overall view demands constant zooming adjustments that can lead to disorientation. 
These findings highlight the need to build on existing AT strategies and provide simultaneous access to the data and its context, which could preserve visual agency and minimize effort. 
This leads to our main research question: How can we provide context while leveraging the advantage of current AT for LVI?

To answer this question, we explored two interaction methods: \dc and \mm (\autoref{fig:system-teaser}). 
\dc is a novel semantic focus+context-inspired approach that creates a compact view of contextual elements around the user's pointer. \mm is an overview+detail-inspired approach adapted for LVI.
Both methods are designed to work in tandem with existing AT to enhance current visual access strategies.
For example, when hovering over a line chart, using \dc, users can simultaneously view both vertical and horizontal values of a point without panning. Similarly, using \mm, users can identify out-of-view peaks or troughs without zoom adjustments. 
We implemented a prototype to test these interaction methods. The scope of the current study allowed us to use manual extraction of chart information; however, using automated extraction~\cite{zheng2025AdvancingChart,ji2025SocraticChart,goswami2025ChartCitorMultiagent} is proposed as a future avenue of research.

We conducted a comparative study with N=22 LVI using the prototype to evaluate the impact of both methods on usability, effort, and access for chart question answering, contrasting them against a baseline of existing strategies. 
To explore integration with existing AT, participants used their preferred AT across all conditions.
Our results show that \dc was significantly more useful than the \bl, reduced effort, and provided significantly more perceived access to charts---self reported---compared to both \mm and the \bl.
\mm was less useful for the task; however, participants recognized its potential to deliver valuable spatial context that aids in orientation and discoverability of chart regions.
Crucially, while \dc increases access, it introduced visual clutter, which particularly hindered participants who experience double vision. 
Based on our findings, we propose design insights for developing tools that support LVI's visual agency by providing seamless access to out-of-view visual context while balancing visual clutter. 
All code and data is made available through the project page\footnote{\url{https://visual-context.github.io/}}.

\medskip

\noindent\textit{Contribution Statement:}
\begin{itemize}
    \item Insights from a \textbf{formative study} with five LVI, exploring how they access charts, challenges they face, what context they use, and differences between chart types. 
    \item \textbf{Two interaction methods}: \dc, a novel semantic focus+context inspired method, and \mm, an overview+detail inspired method, adapted for LVI.  
    \item \textbf{Empirical findings} from a comparative evaluation with 22 LVI using our prototype implementation on an established chart question-answering task with 12 chart types.  
    \item Actionable \textbf{design insights} to develop future AT that provides visual context for LVI in information-dense tasks. 
\end{itemize}

%% file: content/02-background.tex
\section{Related Work}
\label{sec:background}


\subsection{Assistive Tools for Low-vision}
Low vision encompasses a spectrum of impairments that cannot be fully corrected through glasses, medication, or surgery, and may involve reduced acuity, restricted visual fields, light sensitivity, or color vision deficiencies (CVD)~\cite{szpiro2016HowPeople,wang2023UnderstandingHow, 10.1145/3746059.3747704}. To access digital content, LVI rely on AT built into their device or third-party AT~\cite{crossland2014SmartphoneTablet,alcaraz-martinez2024EnhancingStatistical}, which can be broadly grouped into three categories: \textit{size adjustments}, \textit{visual styling}, and \textit{sensory modality transitions}.  These AT are not used exclusively, and LVI often combine them~\cite{szpiro2016HowPeople}.

Size adjustments enlarge on-screen elements via magnifiers~\cite{microsoft2024magnifier,apple2024zoom}, browser zoom~\cite{bigham2014MakingWeb}, or component-level resizing (e.g., icons, text)~\cite{scott2002ImpactGraphical}. Prior works proposed ways to make navigation of magnified visual content less tedious in contexts such as web pages~\cite{billah2018SteeringWheelLocalitypreserving,lee2020TableViewEnabling}, office applications~\cite{lee2021BringingThings}, or presentation videos~\cite{sechayk2025VeasyGuidePersonalized}. 
Visual styling modifies presentation through high-contrast themes, inversion, or aesthetic color filters~\cite{darkreader2014}, which is often combined with personalization~\cite{sechayk2025VeasyGuidePersonalized,zhao2015ForeSeeCustomizable,sackl2020EnsuringAccessibility} to address the wide range of visual conditions. 
Sensory modality transitions focus on non-visual access such as text-to-speech or tactile tools. Text-to-speech ranges from selective text-to-speech~\cite{szpiro2016HowPeople} to full screen readers~\cite{chheda-kothary2023UnderstandingBlind}. Tactile tools are commonly braille displays~\cite{lang2023BrailleBuddyTangible,persinger2025ExploratoryStudy}, and more recently tactile fabrication using 3D printing~\cite{geronazzo2016InteractiveSpatial,reinders2025WhenRefreshable,milallos2024ExploringEffectiveness,gotzelmann2016LucentMaps3D}.

With recent technological advancements---such as AR and eye tracking---some works explore using visual aid augmentations for LVI.
For instance using arrows, outlines, or other visual embellishments in head-mounted AR devices to assist LVI with activities such as running outdoors~\cite{abe2025CanRun}, shopping for groceries~\cite{zhao2016CueSeeExploring}, or interacting with touch panels~\cite{lang2021PressingButton}.
Other works adapted existing paradigms for accessibility to leverage AR, such as screen magnification on real-world objects~\cite{zhao2015ForeSeeCustomizable}.
Using eye tracking for LVI, \citet{wang2024GazePromptEnhancing} developed \textit{GazePrompt}, which uses visual augmentations to enhance the reading experience. 
These works render \textit{visual highlights} to make hard-to-see elements within the user's field of view more accessible to LVI.

Our work extends this approach of visual augmentations; however, instead of highlighting, we superimpose elements outside the field of view of LVI into a compact and consistent layout within their view. To align with how LVI currently use AT, our approach is designed to be combined with existing AT for an improved and personalizable chart reading experience.


\subsection{Focus, Overview, Context, and Details}
There exists a tradeoff between seeing details of an element of interest and maintaining orientation within its broader context~\cite{cockburn2009ReviewOverview+detail,carpendale2001FrameworkUnifying}. Research has explored two main paradigms to balance this tradeoff: overview+detail and focus+context.
Overview+detail interfaces present a detailed view alongside a smaller overview indicating the current focus within the whole~\cite{hornbaek2002NavigationPatterns,baudisch2002KeepingThings}. Because the views are separated, users interact with them independently, though actions in one are typically mirrored in the other. These interfaces are often paired with zooming to support navigation~\cite{burigat2008MapDiagram} and have also been extended to immersive 3D environments~\cite{yang2021EmbodiedNavigation}. One variant is lenses---movable regions that overlay the main view to magnify~\cite{appert2010HighprecisionMagnification} or transform~\cite{tan2004WinCutsManipulating} content---effectively separating overview and detail along the z-axis (depth).

Focus+context interfaces integrate all regions into a single display, often through fisheye views or differential scaling that deliberately distort the presentation~\cite{furnas1986GeneralizedFisheye,sarkar1992GraphicalFisheye,horak2021ResponsiveMatrix}. This approach reduces short-term memory load by preserving both focus and context together~\cite{cockburn2009ReviewOverview+detail}. Alternatives to distortion include translucency blending~\cite{pietriga2008SigmaLenses} or semantic zooming, which adapts presentation based on content type~\cite{agarwal2013WidgetLensSystem}. Other approaches leverage user intent, such as using pointer speed to transition between the context and the focus when exploring maps~\cite{pietriga2008SigmaLenses}, or mapping scrolling speed to change magnification level when reading large documents~\cite{igarashi2000SpeeddependentAutomatic}.

While these paradigms improve navigation, orientation~\cite{hornbaek2002NavigationPatterns,yang2021EmbodiedNavigation}, and exploration~\cite{shamir2012InteractiveVisual} for sighted individuals, little work has examined their impact on LVI~\cite{islam2023SpaceXMag}. For LVI, the challenge of fitting information into limited visual space~\cite{carpendale2001FrameworkUnifying,theofanos2005HelpingLowvision} is compounded by reduced acuity, field of view, or color perception~\cite{szpiro2016HowPeople}. Thus, existing designs may not meet their needs. In this work, we revisit overview+detail and focus+context from the perspective of LVI, exploring necessary adaptations for this demographic.


\subsection{Chart Accessibility for Low-vision}

Making charts accessible remains a significant challenge~\cite{kim2021AccessibleVisualization,elavsky2022HowAccessible}.
Recently, there is an effort to making charts more broadly accessible.
\citet{marriott2021InclusiveData} surfaced gaps in visualization accessibility efforts and highlighted opportunities to elevate access to visualizations---such as the development of new modes of interaction.
\citet{elavsky2022HowAccessible} developed \textit{Chartability}, a set of heuristics for making accessible visualizations. The development of Chartability is deeply grounded in research and extensive real-world experience. They surface many requirements for visual access, such as screen reader or magnification support, sufficient color contrast, appropriate spacing, and preventing obstruction of AT use.
However, accessibility efforts for people with visual impairments have typically centered on non-visual access~\cite{wimer2024VisionImpairments,marriott2021InclusiveData}. Current accessibility approaches include textual descriptions (e.g., longdesc attributes~\cite{w3clongdesc}), inclusion of data tables~\cite{alcaraz-martinez2024EnhancingStatistical,moured2024Chart4BlindIntelligent}, hierarchical navigation via screen readers~\cite{zong2022RichScreen,blanco2022OlliExtensible}, and emerging frameworks such as ARIA-based chart annotations~\cite{w3cariacharts}. 
Other modalities include tactile graphics~\cite{seo2024MAIDRMaking}, haptics~\cite{singh2023GraphoBringing}, and multimodal interfaces (e.g., audio and speech navigation~\cite{zhao2024TADAMaking}). More recently, large language models (LLMs) have been explored as a means for personalized chart access through conversational agents~\cite{kim2023ExploringChart,gorniak2024VizAbilityEnhancing,seo2024MAIDRMeets}. 

Compared with non-visual access, few works have explored how LVI use their residual vision when reading charts.
\citet{prakash2024UnderstandingLow} studied interactions with five types of bar charts through a web-based magnifier, finding that blur and distractor bars led to frequent misjudgments, and that stacked or unaligned bars increased perception errors. They propose future work to explore auto-panning to key chart regions (e.g., axis labels, bar tops). 
\citet{alcaraz-martinez2024EnhancingStatistical} compared WCAG~\cite{wcag21} accessible and non-accessible versions of three web charts and showed that accessible designs improved efficiency, effectiveness, and satisfaction, while identifying remaining barriers such as legend size, tooltip occlusion, and labeling. 
\citet{wang2024HowLowvision} investigated in-the-wild visualization use with AT, documenting how LVI rely on magnification and prior knowledge to piece together partial views, and face color-related challenges even when charts are made accessible. 
Only a handful of works developed AT that target the unique needs of LVI.
Examples include audio-graphic interfaces for math graphs~\cite{dzierzgowska2025AlternativeAudiographic} and smartphone tools such as \textit{GraphLite}, which transforms static bar charts into personalizable interactive views~\cite{prakash2025EnhancingLow}. 

In this work, we contribute to improving chart accessibility for LVI via a novel approach that focuses on providing access to chart context. While access to context is important for interpreting charts~\cite{shneiderman1996EyesHave}, prior work has not explored how context can be integrated into the chart reading experience of LVI.


\subsection{Chart Comprehension}

Successful chart comprehension relies on visualization literacy, defined by \citet[p.~552]{lee2017VLATDevelopment} as \q{[...] the ability and skill to read and interpret visually represented data in and to extract information from data visualizations.} They proposed the Visualization Literacy Assessment Test (VLAT), a standardized test for quantifying users' visualization literacy. It consists of 53 questions across 12 visualization types. Because the full VLAT is fairly long, \citet{pandey2023MiniVLATShort} introduced  \textsc{Mini-VLAT}, a short-form (12-item) alternative that preserves reliability and correlates strongly with the full VLAT.

Research has highlighted both barriers and strategies in how people engage with data displays. \citet{nobre2024reading} examined barriers that limit visualization literacy and emphasized the need for accessible design and instruction.
Other work looked beyond accuracy to study the strategies individuals use in visual problem-solving tasks~\cite{morth2025beyond, lee2015people}. Their findings suggest that adaptive interfaces are needed to accommodate diverse user needs.
Formal instruction also plays a role: studies investigated what students learn in visualization courses~\cite{hedayati2024university}, how cognitive characteristics correlate with visualization literacy~\cite{lee2019correlation}, and how adaptive assessments can better capture individual skill levels~\cite{cui2023adaptive}. These works highlight that visualization proficiency is influenced by individual differences, pointing to the value of personalized educational and assessment strategies.
Related research has compared graph interpretation skills in scientific versus everyday contexts among students~\cite{binali2024high}, while systems such as \textit{Treeducation} teach specific visualization concepts, like treemap layouts~\cite{fuchs2024treeducation}. Additionally, \citet{omelchenko2025CrosslinguisticCultural} demonstrated that culturally adapting the Mini-VLAT for Ukrainian speakers significantly improved assessment accuracy, highlighting the importance of context-specific adaptations.

We use the Mini-VLAT because it (1) provides a community-recognized, standardized baseline for comparability; (2) is psychometrically validated and thus reduces measurement noise; and (3) minimizes participant burden in multi-condition usability studies while allowing us to control for or examine individual differences.

%% file: content/03-formative.tex
\section{Formative Evaluation}
\label{sec:formative}

To understand what LVI seek from tools to support chart reading, we conducted a formative evaluation with five LVI. 
While prior work revealed how LVI experience charts through exploratory inspection~\cite{prakash2024UnderstandingLow,wang2024HowLowvision,alcaraz-martinez2024EnhancingStatistical}, we use Mini-VLAT~\cite{pandey2023MiniVLATShort} for its broader scope of chart coverage~\cite{lee2017VLATDevelopment}. 
Our goal was to understand visual access strategies and AT usage, identify what context LVI use, and any differences between chart types.

\input{tables/03-participants.tex}

\subsection{Participants}
We recruited five LVI participants, two male and three female, ages 28 to 42 (\( M=33.8\), \(SD=5.07\)) through personal connections, see \autoref{tab:formative_participants}. All participants identified as having low vision, with a preference for using their residual vision to access visual media. 
Four participants are active users of screen magnification software on their devices (F1, F2, F4, F5), and three use color filters (F1, F4, F5). F1 and F4 occasionally use a screen reader; however, they did not use it during the study. The study was conducted online via Zoom and lasted around 50 minutes (\(M=47.2\), \(SD=9.1\)). Participants were compensated with 20 USD. 
This study received approval from the Institutional Review Board (IRB) at our institution. Participants provided informed consent and were fully briefed on the research aims and data usage before the study.

\subsection{Procedure}
Upon obtaining consent, we began with pre-test questionnaires covering participant demographics, visual condition, AT usage, and self-reported chart familiarity. 
Before starting the main task, we provided a thorough explanation of the Mini-VLAT's layout, structure, and instructions to ensure a comprehensive understanding. Participants then solved the Mini-VLAT\footnote{\url{https://washuvis.github.io/minivlat/} (Accessed: March 17, 2025)} using their choice of AT.
The procedure concluded with a semi-structured interview, during which participants reflected on their experiences with different charts using the PDF version\footnote{\url{https://washuvis.github.io/minivlat/MiniVlatQUESTIONS_v1.pdf} (Accessed: March 17, 2025)} of the test.

When interacting with and solving Mini-VLAT, we followed the original interface and test conditions, including a randomized question sequence, skip button, and 25-second time limit per question~\cite{pandey2023MiniVLATShort}. 
Participants took a 5-minute break after completing the test before progressing with the study.
Finally, we conducted a semi-structured interview, inquiring about participants' sense of access, strategies, and challenges they faced in accessing the charts.
To reflect on AT usage and different chart types, participants interacted with a PDF containing all charts/questions, simulating access via a different medium. This portion had no time limit, and participants could freely transition between questions.
By uncovering strategies and challenges through observations and feedback, we address the key questions of this study regarding context.

\subsection{Analysis}
Participants shared their screen and video for analysis. F5 experienced technical difficulties and was not able to share their video feed.
The sessions were recorded and transcribed for analysis using OpenAI Whisper~\cite{radford2022whisper} locally, and the first author manually corrected any transcription errors based on the recording. 
The first author viewed each recording multiple times to systematically analyze the screen and video portions. The analysis focused on classifying AT usage and uncovering access patterns. Both transcriptions and viewing notes were analyzed by the first author using reflexive thematic analysis~\cite{braun2019ReflectingReflexive}, where they used their own experiences as LVI as a lens to analyze the data.

\subsection{Findings}
Participants reported diverse familiarity with charts, which is reflected in their performance in Mini-VLAT. Participants F1, F2, and F4 self-reported high familiarity, interacting with them daily, in professional (F2) or educational scenarios (F1, F4). F3 reported medium to high familiarity, and F5 reported limited familiarity, mostly with bar or line charts encountered through news articles.
No chart was correctly solved by all participants, and no participant correctly answered the stacked bar chart. Six chart types---the bar, line, and pie charts, along with the scatterplot, treemap, and histogram---represented the highest success rate, each correctly solved by three participants. 
See \autoref{tab:formative-scores} for all participant performance.

The 25-second time constraint was too short for participants. 
F1 and F5 timed out on eight of the 12 questions, F4 on five, and F2 and F3 on three questions.  All participants expressed frustration with the short time limit. F3 explained, \q{[The time limit] was annoying. It made me feel my disability more. [...] At school I always got extra time, and now I see why.} 
However, while necessary, participants recognized that just providing additional time does not improve access. F4, \q{I could do it with more time, [...] but there would still be a higher visual impact. [...] you're looking at a lot of different areas of the screen at the same time, aren't you? It makes it very difficult.}
Beyond time constraints, we detail participants' method of access and AT use, as well as several visual challenges they experienced.

\input{tables/03-quiz-results.tex}

\subsubsection{Current AT preferences}
Participants with low visual acuity used magnification (F1, F2, F4, F5), with F1 using a docked magnifier panel, and F2, F4, and F5 using full-screen magnification. F1 and F4 combined magnification with browser-based page-level zoom; however, they did not alter the page zoom during the study. 
F2 used a \textit{dark mode} browser extension that inverted the interface colors, but turned it off midway through the test due to issues with contrast.
F3 did not use any AT during the study. 
All participants who use AT commented that AT provided significant benefits, such as dynamic magnification (F4, F5) and color filters (F2). 
Reflecting via the PDF, F2 commented, \q{I need a giant piece of paper that is unwieldy, and very hard to deal with. [...] Thankfully, these days I can do everything digital.} 
Similarly, F4 expressed, \q{I like the computer because I can make it as big as I need it to be, and particularly, because my vision fluctuates, I can have it smaller at times that I'm doing OK and bigger when I'm a bit tired.}

\subsubsection{Strategies when accessing charts}
Most participants explored the chart's layout to answer the questions, using a similar structured approach that progressed from the title through to the axes or legend, and then to the data. 
However, F5 preferred to exercise a more exploratory approach on some of the questions, \q{I wanted to get the general information and then look at the answers, so I didn't have time to find the actual answer. I started by trying to understand the trends in the chart but sometimes the questions are not related, but I want to understand the overall picture at first.}

The chosen strategy was also influenced by the question or chart type and their familiarity with it. For charts with a legend, participants commonly viewed the legend before the axes. For unfamiliar or complex charts (such as the bubble chart), participants explored each layout element more than once before trying to answer the question. For questions asking about trends (e.g., \q{most common trip distance}), the strategy changed accordingly. 
Additionally, similarity between charts can lead to confusion and affect the ability to choose the correct strategy. F1 explained, \q{Histograms are challenging for me because they are so visually similar to bar charts. Sometimes it would take time to figure out what type of chart I was looking at and what type of information I would need to be looking for.}

\subsubsection{Visual navigation challenges}
Participants reported common challenges such as low-contrast colors and low-resolution images when magnified~\cite{alcaraz-martinez2024EnhancingStatistical}, and visual navigation challenges due to viewing charts in fragments. We outline four themes.

\paragraph{\textbf{Excessive need for manual operations.}}
Participants cannot view the complete chart and its details simultaneously, and must pan between fragments using the mouse pointer (F1, F2, F4, F5) or physically (F3). 
F3 shared, \q{With my field of vision, I need to focus on different sections, so I cannot see everything together, so I need to navigate physically and look around the plot.} 
When transitioning between fragments, participants use the complete view of a chart for orientation. 
F2 explained, \q{My process is big picture first, point of interest, zoom in to read. And then, either I already know where my next point of interest is, and I can just pan towards it, or zoom out to find it and then zoom back in.}

\paragraph{\textbf{Inconsistent layout.}}
Participants had expectations for the position of elements based on their experience with charts. 
When axes are present, participants scanned from the bottom-left placement, which they expect to be the start.
Inconsistent positions---such as with the legend---led to wasted effort or missed information.
F4 explained, \q{[The legend] wasn't always where I was expecting it to be. And sometimes, for a couple of them, I didn't even realize that there was a legend.}
F3 noted, \q{Consistent position of the legend is much better, [...] I do not need to look for 'where to see' so much.}

\paragraph{\textbf{Spread-out context}}
For participants, spread-out context makes constructing mental maps challenging.
F2 explained, \q{If I'm looking at a chart that has many kinds of memory allocation operations, I cannot take it all in at once, I have to refer to the legend and axis and so on many times.}
One chart, showing the number of Olympic medals, had the legend embedded in the chart design. 
F3 expressed, \q{It was nice that the bar colors matched the medal color, I didn't need to look at the legend.}
Most participants (F2, F3, F4, F5) expressed desire to have the context placed near or embedded in the data. 
F2 highlighted this need, \q{You have a little legend here, and then you have to go [to the data], and right now, I have neither of the axes in view, and I've already forgotten the legend. [...] I have no context for what I'm looking at.}

\paragraph{\textbf{Tracking and comparing.}}
Most charts required participants to track data points to axis values or make comparisons between points (e.g., identifying the maximum or minimum).
On tracking, F2 explained, \q{I had to focus on the number down, then pan up to the height of the bar, then back down to make sure I was reading the distance correctly, and so on. I cannot read these things at the same time.}
Tracking becomes more difficult with magnification, as F4 noted, \q{You're having to follow the lines along the screen with your mouse for much further, because you're zoomed in much further, so, you have to follow the line carefully for longer to get to the information.}
Similarly, comparing the data was challenging, with participants (F1, F2, F4) suggesting the use of a ruler as a visual anchor. 
F4 commented, \q{It would be helpful, especially for bar charts, to have a ruler so it would be easier to follow.}
While F3 used the chart grid lines as rulers for tracking and comparing, most participants did not notice them due to the low contrast, resorting to other methods.
F1, who used a docked magnifier, used the boundary of the dock as a makeshift ruler to compare bars. 

\subsection{Design Goals}
\label{sec:design-goals}
Our findings confirm that LVI experience charts in fragments, with significant effort dedicated to panning, zooming, and constructing mental maps of the information~\cite{wang2024HowLowvision,alcarazmartinez2022MethodologyHeuristic}. 
In charts, LVI use the axes and legend to understand the data, the \textit{big picture} to orient within the data, and the grid lines as visual anchors for tracking or comparing data.
Therefore, having access to the context and the details at the same time is essential to improve visual access. Based on these findings, we outline design goals (DGs) for AT to support LVI in chart reading.

\begin{itemize}[leftmargin=*]
    \item \textbf{DG1:} Make chart context accessible: overview, legend, axes, and grid lines. 
    \item \textbf{DG2:} Reduce effort required for visual navigation.
    \item \textbf{DG3:} Support different chart reading strategies.
    \item \textbf{DG4:} Be flexible for fluctuating LVI needs.
\end{itemize}

When reading charts, LVI spend significant effort (1) adjusting zoom levels to access the overview, (2) pan around to access the legend and axes, and (3) tracking and comparing information with no access to grid lines (DG1). 
If chart layouts are not consistent, LVI require additional effort to scan for context (e.g., legend), and compact layouts where the context is near the data would further reduce required effort (DG2).
LVI have different preferred strategies for accessing charts, which can be influenced by different chart types and chart familiarity (DG3).
Finally, LVI have diverse and fluctuating access needs that cannot be addressed with a one-size-fits-all approach (DG4).

%% file: tables/03-participants.tex
\begin{table*}[t]
\caption{Formative Study Participant Demographics. \textbf{A/G}: Age and Gender. \textbf{CVD}: Color Vision Deficiency. \textbf{O}: Onset (A=Acquired, C=Congenital). \textbf{LB}: Recognized as Legally Blind. \textbf{MS}: Monitor Size. \textbf{CF}: Charts Familiarity. \textbf{Gender}: M (Male), F (Female).}
\label{tab:formative_participants}
    \centering
\begin{tabular}{l l l c l c c c}
    \toprule
    \textbf{PID} & \textbf{A/G} & \textbf{Diagnosed Condition} & \textbf{O} & \textbf{CVD} & \textbf{LB} & \textbf{MS} & \textbf{CF}\\
    \midrule
    F1 & 28/F & Strabismus, Chiari malformation & C & Reduced Contrast & \textbf{T} & 23" & 5\\
    F2 & 32/M & Bilateral juvenile glaucoma, cataracts & C & --- & \textbf{T} & 32" & 5\\
    F3 & 30/F & Retinitis Pigmentosa, night blindness & C & --- & \textbf{T} & 27" & 4\\
    F4 & 42/F & Albinism, nystagmus, astigmatism & C & Light Sens. & \textbf{T} & 27" & 5\\
    F5 & 37/M & Albinism, nystagmus, astigmatism & C & Light Sens. & \textbf{T} & 16" & 3\\
    \bottomrule
\end{tabular}
\end{table*}


%% file: tables/03-quiz-results.tex
\begin{table}[h]
    \caption{Mini-VLAT results for formative study participants (\cmark=correct, \xmark=incorrect, TO=timeout, SK=skipped). Each participant had identical 12 chart-question pairs in randomized order. For score calculation, we followed \citet{pandey2023MiniVLATShort}.}
    \label{tab:formative-scores}
    \centering
    \begin{tabular}{ccccc|l}
    \toprule
         \textbf{PID}&  \cmark & \xmark & \textbf{TO} & \textbf{SK} & \textbf{Score}\\
         \midrule
         F1& 
     4& --& 8&--&4\\
 F2& 8& 1& 3&--&7.75\\
 F3& 7& 2& 3&--&6.5\\
 F4& 6& 1& 5&--&5.75\\
 F5& 1& 3& 8&--&0.25\\
 \bottomrule
 \end{tabular}
\end{table}

%% file: content/04-implementation.tex
\section{Interaction Methods}
\label{sec:implementation}
Based on our DGs (\S\ref{sec:design-goals}), we developed two interaction method prototypes: \dc and \mm. To make the axes, legend, grid lines and the overview accessible (DG1), we project them into a compact and consistent area centered around the pointer (DG2). Both methods integrate into existing AT to support different access strategies (DG3), and incorporate personalization settings to support fluctuating user needs (DG3). In this section, we explain the prototype implementation of the two interaction methods.

The prototype implementation was developed as a web-based interface using React~\cite{react} and shows context visually, localized around the user's pointer inside the so-called Overview Area (OA).
The current implementation uses rasterized bitmap charts with bounding-box annotations of chart context (legend and axes). The annotations are currently added manually by the first author. This manual process can be automated by using existing analysis methods~\cite{zheng2025AdvancingChart,ji2025SocraticChart,goswami2025ChartCitorMultiagent}; however, since these methods are not perfect, it is left as future work.
Both interaction methods include personalization settings, following existing practices on AT for LVI~\cite{sechayk2025VeasyGuidePersonalized,zhao2015ForeSeeCustomizable,prakash2025EnhancingLow,billah2018SteeringWheelLocalitypreserving}. 
All setting changes are visible immediately, following the WYSIWYG\footnote{``What You See Is What You Get''~\cite{2025Wysiwyg}} design approach~\cite{billah2018SteeringWheelLocalitypreserving}.

\begin{figure}[htbp]
    \centering
    \includegraphics[width=\linewidth]{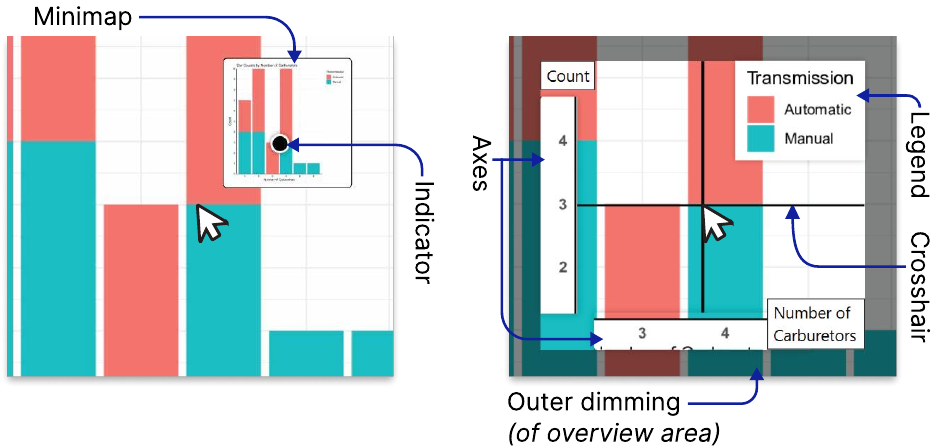}
    \caption{\mm (left) and \dc (right) with each of their components.}
    \label{fig:interface-overview}
    \Description{Visual overview of the two interaction methods. Mini-map includes an indication for the small map itself and the pointer indicator. Dynamic context includes the overview area with outer dimming, projected axes and legend, and the crosshair at the center with its lines extended to the ends of the area.}
\end{figure}

\subsection{\dc}
\dc is a novel semantic interaction method inspired by focus+context approaches. We project contextual information onto a compact area around the user's pointer (\autoref{fig:interface-overview}). The information appears when the user hovers over a chart, and includes three core contextual elements: axes, legend, and grid lines. 
For context updates, we listen to pointer movement changes. On position updates, \dc displays the context inside the OA using projected chart portions, based on bounding-box annotations and OA dimensions.
For instance, if the user points to the top of a bar which indicates that there are three cars (Y-axis) with five carburetors (X-axis), the bottom of the \dc area will show the portion of the X-axis centered around the value \verb|5|, and the left of the \dc area will show the portion of the Y-axis centered around the value \verb|3|. 

To preserve the visual integrity of the charts, we use projection of axes and legend to the OA. 
While reconstructing the context inside the OA could provide other accessibility improvements (e.g., font size), we chose this approach to preserve the chart creator's intent, since reconstruction is not always reliable~\cite{prakash2025EnhancingLow, zheng2025AdvancingChart}.
The X-axis is positioned at the bottom of the OA, and the Y-axis is flushed to the left edge. The titles of the axes are positioned on the bottom-right and top-left of the overlay area, respectively. 
The legend is positioned at the top-right corner of the OA. 
Both the axes and legend preserve their original dimensions in the chart when projected.
To provide the grid line context, we implement a crosshair-like element positioned at the center of the OA. 

\begin{figure}[htbp]
    \centering
    \includegraphics[width=0.95\linewidth]{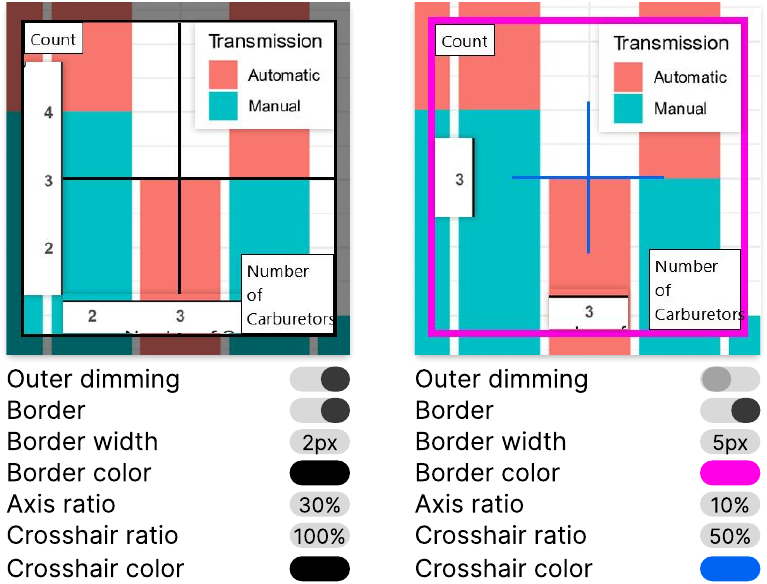}
    \caption{Default configuration of \dc (left), and an alternative configuration (right) using some of the available settings.}
    \label{fig:local-focus-examples}
    \Description{Visual examples from the interface of the system. The default settings have outer dimming on, with a 2-pixel black border, axis ratio of 30 percent, and a black crosshair that reaches to the edges of the overlay area. The alternative settings have outer dimming off, a 5-pixel magenta border, axis ratio of 10 percent, and a blue crosshair that reaches halfway to the edges of the overlay area.}
\end{figure}

\subsubsection{User customization}
\begin{figure}[htbp]
    \centering
    \includegraphics[width=0.9\linewidth]{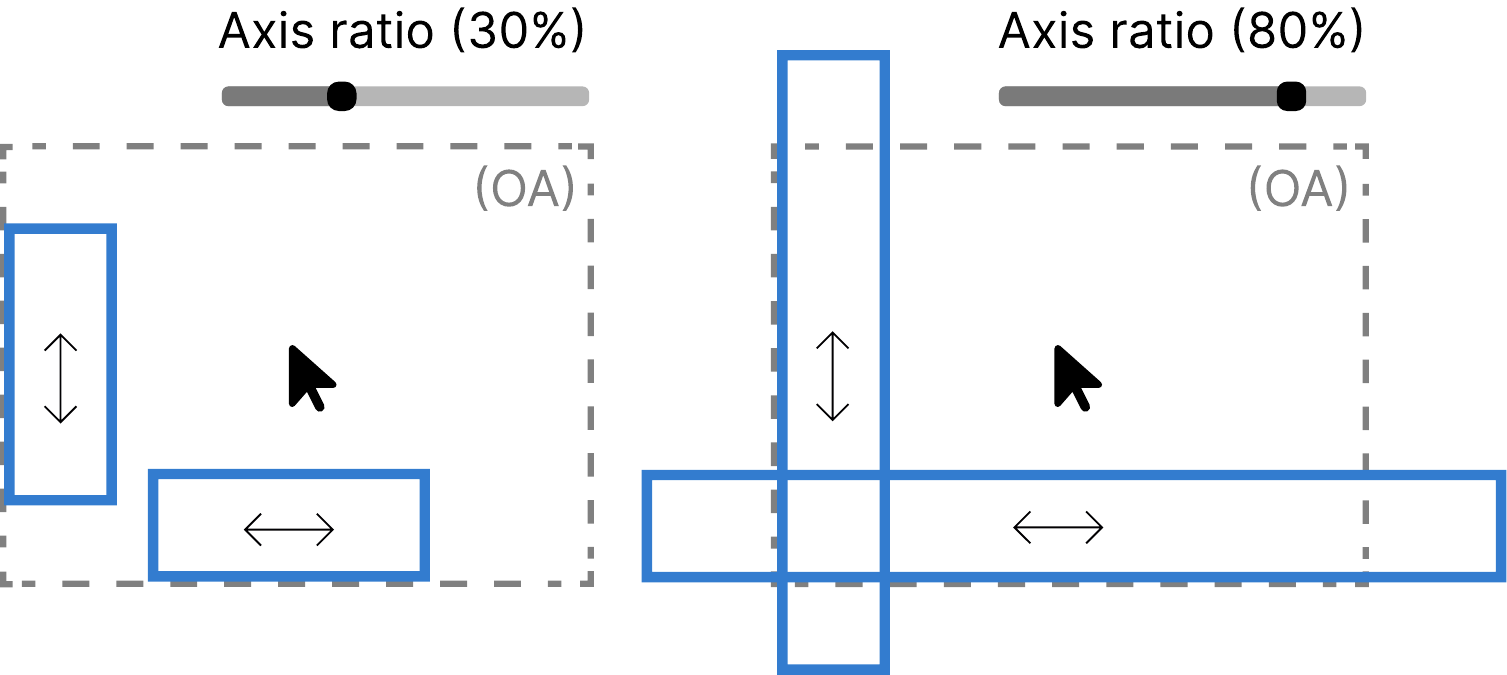}
    \caption{Examples of different axis ratio configurations. (left) value of 30\%, (right) value of 80\%. When axes are larger than the overlay area, they intersect at its bottom-left corner.}
    \label{fig:axis-ratio-examples}
    \Description{Visual that shows how the axis ratio affects the projection of the axes in the overlay area.}
\end{figure}

We provide configurable settings in a dedicated settings window or through user interactions. 
See \autoref{fig:local-focus-examples} for a view of the default and an alternative configurations of \dc. 
First, the OA dimensions (width and height) can be adjusted. Additionally, users can change its border, thickness and color, as well as control an outer dimming functionality, which darkens all visual content outside the area---mitigating visual clutter. The context inside the OA is enabled by default and can be toggled on and off via a left-click of the pointer.
The projected axes can be adjusted through an \textit{axis ratio} slider, which controls the percentage of the axis that is projected into the overlay area, centered around the user's pointer. For example, given that $A_r=0.3$ is the current axis ratio value, at every point, 30\% of an axis will be projected such that 15\% will be on one side of the pointer and 15\% on the other side. When $A_r=1$, the entire axis will be visible at all times, and both axes will intersect at the bottom-left corner of the overlay area (\autoref{fig:axis-ratio-examples}).
For the grid lines, the thickness, color, and opacity can be controlled, along with the ratio of the lines in relation to the overlay area---how far they reach from the pointer position. For example, when set to 50\%, the vertical line would reach the halfway mark on the top/bottom and left/right sides of the OA. When set to 100\%, the lines would reach the edges of the OA.

\subsection{\mm}
\mm is an adaptation of overview+detail to LVI, with the goal of providing the full view of a chart when focusing on the details. 
To provide an overview-type context, we position a minified chart view within the OA (\autoref{fig:interface-overview}). On the minified view, we indicate the current pointer position in relation to the chart. 
\mm contains a circular indicator that points to the current location of the user's pointer on the chart---common in overview+detail interfaces. 
The indicator moves with the pointer, similar to how context updates in \dc.

\begin{figure}[htbp]
    \centering
    \includegraphics[width=0.95\linewidth]{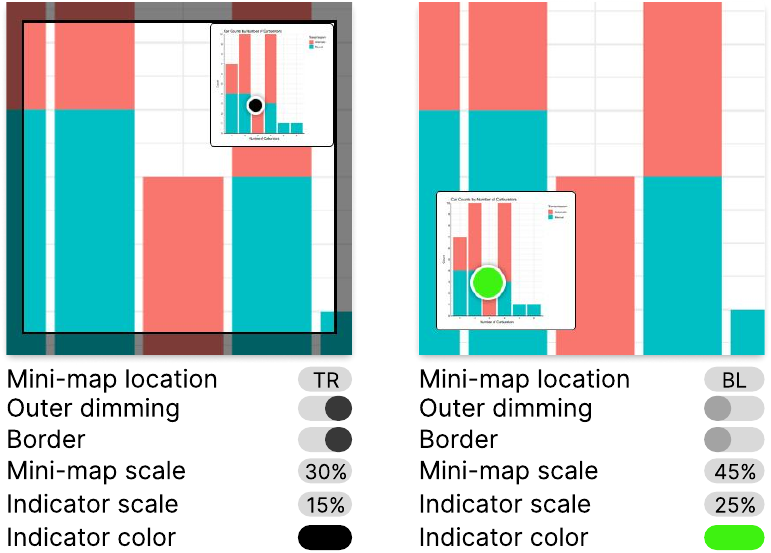}
    \caption{Default configuration of \mm (left), and an alternative configuration (right) using some of the available settings.}
    \label{fig:mini-map-examples}
    \Description{Visual examples from the interface of the system. The default settings have outer dimming on, with a 2-pixel black border, scale of 30 percent for the mini-map, positioned at the top-right corner, and a black pointer indicator with a scale of 15 percent of the mini-map. The alternative settings have outer dimming and border off, scale of 45 percent for the mini-map, positioned at the bottom-left corner, and a green pointer indicator with a scale of 25 percent of the mini-map.}
\end{figure}

\subsubsection{User customization}
\mm shares the same OA configuration as \dc (dimension, border, and outer dimming), with some additional settings.
See \autoref{fig:mini-map-examples} for a view of the default and an alternative configurations of \mm. 
Similar to \dc, the visuals of \mm are on by default and can be toggled off and on via the left click.
The minified view can have its \textit{position} adjusted to be flush with any of the four OA corners (top-left, top-right, bottom-left, bottom-right), and its \textit{size} adjusted in relation to the width of the OA. 
The position indicator is visualized as a circle with a white border and a shadow effect, and can have its \textit{size} and \textit{fill color} changed.

%% file: content/05-user-study.tex
\section{User Study}
\label{sec:user-study}

To evaluate our proposed interaction methods (\S\ref{sec:implementation}), we conducted a comparative study with three conditions: \textit{\bl}, \textit{\dc}, and \textit{\mm}. To maintain external validity and explore how our interaction prototypes integrate with existing AT, participants could choose to use any AT in all conditions. 
Few studies have examined overview+detail and focus+context approaches for LVI in navigating complex visual interfaces~\cite{billah2018SteeringWheelLocalitypreserving}. Thus, we designed this study to compare these approaches through our proposed interaction methods. 
This study was guided by the following research questions:  
\begin{itemize}[leftmargin=*]
    \item \textbf{RQ1:} Does \mm or \dc improve efficiency and accuracy?  
    \item \textbf{RQ2:} How will \mm or \dc influence AT usage?
    \item \textbf{RQ3:} Will chart reading strategies change with \mm or \dc?
    \item \textbf{RQ4:} How does \mm or \dc affect usability and workload?
    \item \textbf{RQ5:} How does \mm or \dc influence access to charts?
\end{itemize}

Our RQs are motivated based on several hypotheses. We hypothesize that having access to the context using a compact layout that follows the pointer will improve the efficiency of accessing charts and the ability to answer the test questions accurately (RQ1). As we designed our prototype interactions with integration into AT in mind, we hypothesize that they will not change how participants use AT (RQ2). Chart context is placed around the pointer; therefore, we hypothesize that the way LVI access charts will change, with fewer navigation operations towards the axes or legend (RQ3). Similarly, we hypothesize that the interaction methods will reduce the workload and improve the usability for chart question-answering (RQ4). Finally, we hypothesize that a compact, consistent, and customizable layout will increase the sense of access to charts (RQ5).

\subsection{Participants}
We recruited 22 LVI (\autoref{tab:study_participants}) through public posts in blind and low-vision organizations, closed communities, and snowball sampling. Participants’ ages ranged from 23 to 66 ($M=41.4$, $SD=14.2$), with 12 male, 8 female, 1 agender, and 1 non-binary.  
Eligibility criteria required participants to be at least 18 years old, have low vision, and use residual vision to access digital content. 
Four participants did not identify as legally blind. 
During the study, 14 participants used a screen magnifier, 10 changed the browser zoom level, 3 used pinch-to-zoom, 5 used full-screen color filters, 3 used color themes (\eg high-contrast), and 5 used a customized pointer. For a detailed list of AT usage, see \A\ref{apx:study-at-use}.
Sessions lasted around 90 minutes ($M=91.7$, $SD=21.9$), and participants were compensated \$30. 
This study received approval from the IRB at our institution. Participants were fully briefed on the research aims and data usage and provided informed consent before beginning the study.

\input{tables/05-participant-table}

\subsection{Material}
As in the formative study (\S\ref{sec:formative}), we used the Mini-VLAT~\cite{pandey2023MiniVLATShort} to evaluate our prototype interaction methods across 12 common chart types.  
For this within-subject evaluation with three conditions (\bl, \mm, and \dc), we created two additional variations ($v1$ and $v2$) to the original Mini-VLAT chart-question set ($v0$). Variations were introduced in two primary ways: changes to chart data (permutation, added noise, or magnitude) and changes to questions (orientation or topic). All variations were iteratively reviewed by two visualization experts, with adjustments made based on their feedback. Throughout this process, we prioritized preserving the semantics of each question-chart pair and maintaining comparable difficulty (\A\ref{apx:vlat-variants}).  
We generated an additional set of three question-chart pairs (pie chart, line chart, stacked bar chart) for the tutorial in each condition. These example questions were derived from the \verb|mtcars| dataset (\A\ref{apx:example-charts}). No participant measurements were collected from these example questions, and the same set was used across all conditions.

\begin{figure}
    \centering
    \includegraphics[width=\linewidth]{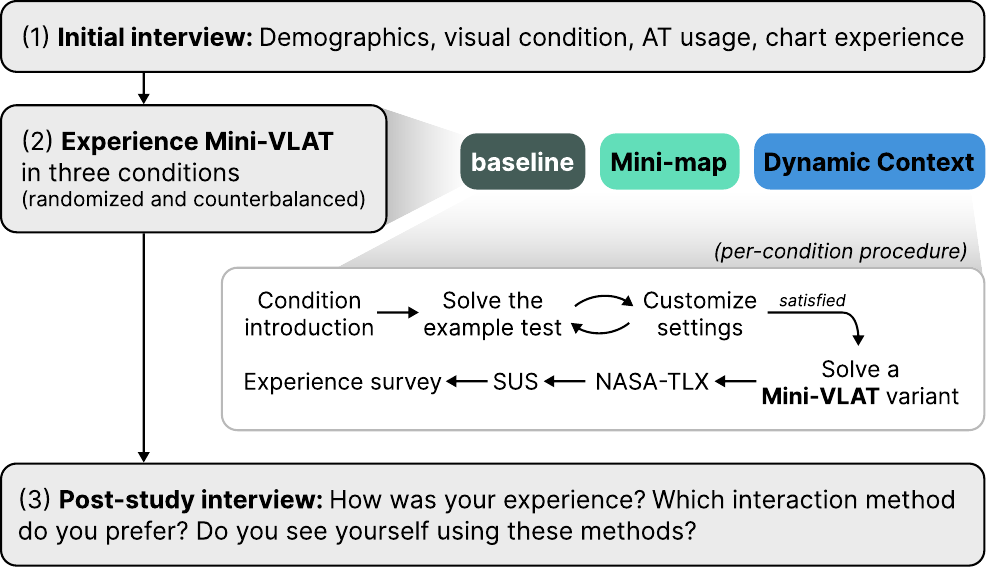}
    \caption{Outline of the study procedure. Participants repeat the process of viewing the example questions and adjusting their settings until they are satisfied.}
    \label{fig:study-procedure}
    \Description{Flow diagram of study procedure based on the provided description. From the initial interview, the three conditions (baseline, \mm, \dc), to the exit interview.}
\end{figure}

\subsection{Procedure}
The study consists of three parts: the initial interview, experiencing Mini-VLAT for the three conditions, and a post-study interview (\autoref{fig:study-procedure}).
We started by explaining the study procedure and which data was collected.
Upon consent, we conducted an initial interview to collect participants’ demographic information, visual conditions, use of AT, and familiarity with charts (self-reported). See \A\ref{apx:initial-questions} for the full list of questions.
We introduced the interface of the study apparatus, then continued to the conditions, the order of which was counterbalanced using the Latin Square method and randomly distributed among participants (\A\ref{apx:counterbalance}). 
Mini-VLAT variations were distributed so that each appeared in every condition exactly twice, ensuring full coverage.

At the start of each condition, we introduced the relevant interaction method and instructed participants to experience the example test, which uses three questions to simulate Mini-VLAT. 
After completing the example test, participants navigated to the customization page.
On this page, we explained the available settings and addressed any questions. Participants could change any setting, with the sample chart on the configuration page updating in real-time. Beyond the sample, participants could go back and forth between the configuration page and the example test to verify their selected configuration.
Once participants expressed satisfaction, they could progress to solving the assigned Mini-VLAT variant.

For both the example and the Mini-VLAT variant, we followed practices for LVI in educational settings~\cite{kellaghan2019PublicExaminations} and set the time limit to 150\% of the original time limit~\cite{pandey2023MiniVLATShort}, resulting in a 38-second limit per question. 
Participants were instructed to avoid random guesses and skip questions if unsure. When the time limit was exceeded, a message prompted them to move to the next question.
After completing the test, participants completed three sets of subjective evaluation questionnaires: task workload (based on NASA-TLX~\cite{hart1988DevelopmentNASATLX}), the System Usability Scale (SUS)~\cite{brooke1996SUSQuick}, and experience-based questions. All questionnaires used a 5-point Likert scale. 
Each condition ended with a two-minute break to minimize eye strain and fatigue, following suggestions from prior work~\cite{sechayk2025VeasyGuidePersonalized}.
After completing all three conditions, we concluded with a semi-structured interview in which participants reflected on their experiences. See \A\ref{apx:post-study-questions} for the full list of questions of the post-study interview.

\subsection{Measurements and Data Analysis}

\subsubsection{Interaction data}
We analyzed participants' interactions with the charts via pointer movements across all conditions. Pointer positions were recorded at a 30 Hz sampling rate. Each sample is represented as a tuple $(x, y, t)$, where $x$ and $y$ are the normalized horizontal and vertical positions, and $t$ is the timestamp. Using this data, we (1) visualized the trajectory and density of chart interactions, and (2) compared trajectories across conditions for each chart.
For trajectory comparisons, we resampled each trajectory to 500 steps, computed pairwise distances using Dynamic Time Warping~\cite{sakoe2003dynamic}, and applied PERMANOVA~\cite{anderson2001new} with 999 permutations to test for \textit{Condition} effects. For charts with significant results, post-hoc pairwise PERMANOVAs with Holm’s correction for multiple comparisons~\cite{holm1979simple} were conducted.

\subsubsection{Test performance}
We evaluated participants' performance across all tests and conditions using two measures: (1) \textit{test score}, and (2) \textit{test duration}, defined as the total test time.  
To compute test scores across conditions, we followed standard scoring guidelines~\cite{pandey2023MiniVLATShort}: correct answers received $+1$, timeouts/skips $0$, and incorrect answers $-1/N_{\text{op}}$, where $N_{\text{op}}$ is the number of answer options.  
To calculate the test time and avoid influence from slow connectivity, we summed the answering time (up to timeout) for all questions. 

To account for variability in participants' visual conditions, we used mixed-effects analysis. The within-subject factor \textbf{Condition} had three levels (\bl, \mm, \dc), and to validate counterbalancing, we included a between-subject factor \textbf{Order}. Normality was assessed using the Shapiro–Wilk test.  
For normally distributed measures, we fit Linear Mixed-Effects Models (LME), performed ANOVA~\cite{kuznetsova2017lmertest}, and conducted Tukey’s HSD post-hoc comparisons. For non-normal measures, we applied Aligned Rank Transform (ART) ANOVA~\cite{wobbrock2011aligned} with ART contrasts~\cite{elkin2021aligned}, using Tukey’s HSD adjustments. Effect sizes are reported as partial eta squared ($\eta^2_p$), with 0.01, 0.06, and 0.14 indicating small, medium, and large effects~\cite{cohen2013statistical}.

\subsubsection{Subjective metrics}
We assessed participants' experiences using three questionnaires: (1) a task-load questionnaire based on NASA-TLX~\cite{hart1988DevelopmentNASATLX}, (2) the SUS for usability, and (3) experience-based questions inspired by prior work~\cite{billah2018SteeringWheelLocalitypreserving,sechayk2025VeasyGuidePersonalized}.  
All questions used a 5-point Likert scale: task-load responses ranged from \textit{very low (1)} to \textit{very high (5)}, while SUS and experience responses ranged from \textit{strongly disagree (1)} to \textit{strongly agree (5)}. Task-load and experience questions were analyzed individually, whereas SUS responses were first aggregated into a SUS score prior to analysis. The analysis procedures followed the approach described in the previous section.

\subsubsection{Qualitative data}
For qualitative analysis, we used participants’ video, audio, and screen recordings. 
Two participants (P2, P4) lacked webcam access, and one (P6) had technical issues preventing video recording. All sessions were recorded and transcribed locally with OpenAI Whisper~\cite{radford2022whisper}. 
The first author reviewed each transcript alongside the recordings and manually corrected any transcription errors. 
As in \S\ref{sec:formative}, we watched all recordings multiple times, taking notes on participants’ physical movement patterns, access strategies, and AT use. 
We used reflexive thematic analysis~\cite{braun2019ReflectingReflexive}, foregrounding the first author’s perspective as an individual with low vision. Coding and initial theme development were led by the first author, who maintained reflexive memos to document positionality and interpretive decisions. In line with \citet{nowell2017ThematicAnalysis}, we conducted meetings with co-authors to provide opportunities for critical discussion and refinement of interpretations.

\subsection{Results}
\label{sec:results}

\subsubsection{Test efficiency and accuracy (RQ1)}
All conditions were statistically indistinguishable for the scores, completion times, and number of timed-out questions. For completeness, we report the descriptive statistics. 
Participants scored on average $7.39$ ($SD = 2.14$) with \bl, $6.52$ ($SD = 2.74$) with \mm, and $6.23$ ($SD = 2.77$) with \dc. 
Completion times were on average $253.08$\,s ($SD = 50.43$) for \bl, $291.72$\,s ($SD = 63.88$) for \mm, and $287.58$\,s ($SD = 67.80$) for \dc. 
The number of time-outs on average were $0.95$ ($SD = 1.21$) with \bl, $1.64$ ($SD = 2.01$) with \mm, and $1.59$ ($SD = 1.92$) with \dc.See \autoref{tab:test_statistics} for more details.

\input{tables/05-test-statistics}

\subsubsection{Influence on AT usage (RQ2)}
Across all conditions, participants used a variety of AT (see \A\ref{apx:study-at-use}). In most cases, both \mm and \dc were successfully used together with AT; however, this was not always the case. 
For both interaction methods, the most common integration challenge was with \textbf{color filters and themes}. In the web browser, under high-contrast themes, many color settings were overridden by the theme. For instance, P21 could not use outer dimming or modify any of the colors. 
Similarly, using a color inversion filter, the outer dimming brightened surroundings rather than darkening them, forcing participants with light sensitivity to disable it. 
Participants who switched between color modes found that configurations sometimes worked in one mode but not the other (\eg P5, P10, P11).

Not all \textbf{magnification methods} integrated smoothly with \dc and \mm. Participants used various magnification methods, including full-screen, lens, browser-based zoom changes, and pinch-to-zoom. Some screen-magnifier users struggled to adjust the OA due to high magnification (P3) or because their magnifier allowed the pointer to move only within the magnified region (\eg P7, P8, P19), effectively clipping the displayed context of \dc and \mm near display edges. For pinch-to-zoom users, technical limitations caused both \dc and \mm to have inconsistent sizes, since the OA could not adapt dynamically to pinch-to-zoom changes.

Several participants (P1, P2, P5, P6) expected \dc to replace their magnification AT, streamlining the interaction. As P1 noted, \q{It would have been nice to have a magnifier option on the tool, like a spot magnifier, because I find that using the browser magnification applies for everything, and the computer magnifier is a lot sometimes, and can make it very tedious to navigate.} Finally, this expectation led some (\eg P9) to perceive the context in \dc as smaller than it was, even though the scale is preserved.

\subsubsection{Strategies for chart reading (RQ3)}
\begin{figure}[htbp]
    \centering
    \includegraphics[width=\linewidth]{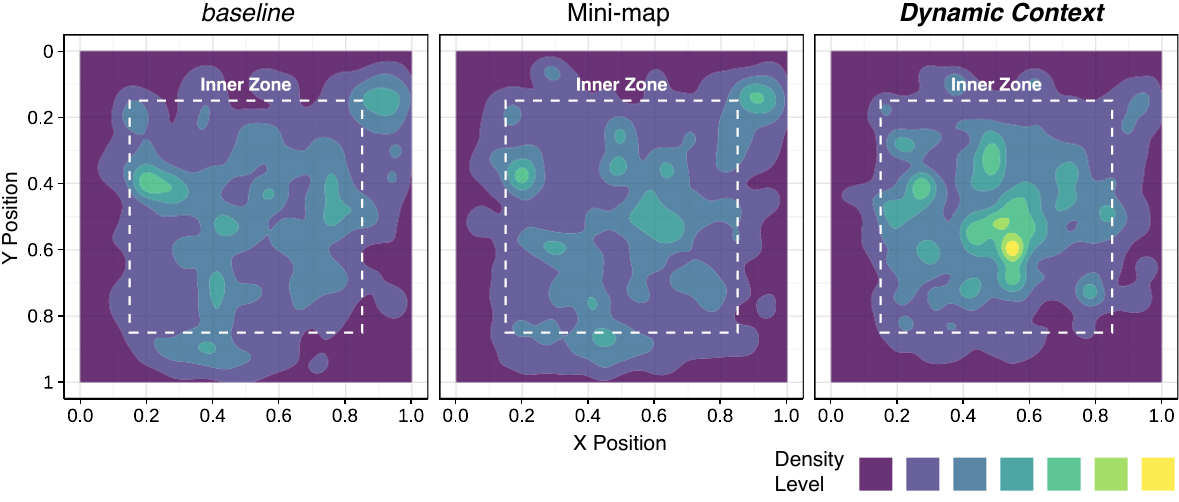}
    \caption{Density of pointer movement across conditions. With \dc, pointer activity appears more centralized compared to \bl and \mm. The inner zone marks the central 70\% both horizontally and vertically, where data is likely to be positioned.}
    \label{fig:interaction_density}
    \Description{Density of pointer positions across all conditions. In the Dynamic Context condition, pointer activity is more centralized compared to baseline and Mini-map. The inner zone marks the central 70\% both horizontally and vertically, representing the main portion of the chart and excluding contextual elements.}
\end{figure}

As in \S\ref{sec:formative}, most participants used a structured approach. Since each chart was paired with a question, they typically read the question first and then inspected the chart, producing concentrated activity in the top-right area adjacent to the question (\autoref{fig:interaction_density}). In the \bl condition, participants usually navigated to one of the axes, identified the relevant value, moved perpendicularly towards the location of interest, and then navigated to read the corresponding value on the other axis. When a legend was present, they often consulted it before the axes.
To track and compare values, participants used additional methods beyond panning. Participants used horizontal/vertical scrolling (\eg P2, P6, P9, P12, P13, P21), their finger (\eg P14, P19), or even aligned the relevant area with the vertical/horizontal edge of their display (\eg P3, P12). Scrolling delivered stable panning with easier control, and using the finger or the display edges provided a clear visual anchor.

\begin{figure}[ht]
    \centering
    \includegraphics[width=0.95\linewidth]{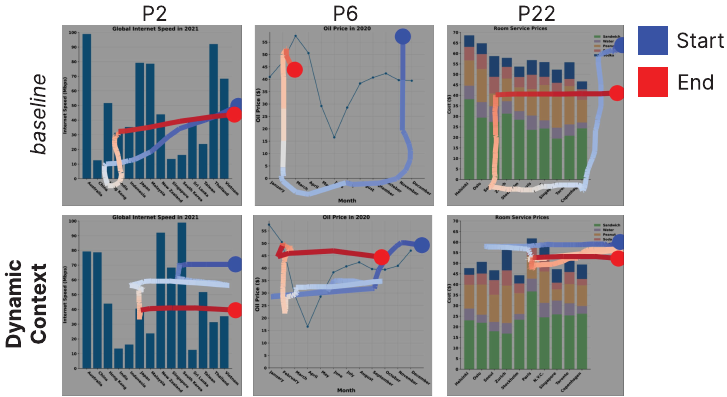}
    \caption{Examples from P2, P6, and P22 comparing chart access in \bl versus \dc using time-aligned scanpath visualization.}
    \label{fig:context_examples}
    \Description{Examples from P2, P6, and P22 comparing chart access in baseline versus Dynamic Context using time-aligned scanpath visualization. In the Dynamic Context mode, the scanpaths were more direct.}
\end{figure}

\paragraph{\textbf{Using \dc (focus+context).}}
\dc had an impact on participants' access strategies. Several participants (\eg P2, P6, P12, P13, P18, P22) adapted their approach when using \dc (\autoref{fig:context_examples}). With \dc, they panned toward the chart center, used the projected axes to target the relevant value, and adjusted their trajectory accordingly. Once at the area of interest, they shifted their gaze to the other axis to read the required value, using the crosshair as a visual anchor. This strategy concentrated pointer interactions around the chart center (\autoref{fig:interaction_density}).
Some participants instinctively kept panning toward one of the axes even with \dc (P5, P6, P19). In these cases, they still benefited from the projected axes, effectively skipping the second panning stage.  
Some participants adapted their strategy during the test (\eg P5, P6, P13). P6 commented, \q{At first I noticed I would go to the axes as I am used to, but then I realized I don’t need to.} P19 explained, \q{In the first couple of questions I was mostly just confirming that the proportions matched even when I zoomed in.} Similarly, P5 noticed the legend in the \dc view and remarked, \q{I liked that the key, I discovered that it was also brought into view. I was working on setting it up to where it would be close so I could try and compare the colors.}
For some charts, the displayed context was too narrow. For the area chart (Q10), the X-axis displayed \textit{April}, \textit{July}, and \textit{October} for two consecutive years; however, both years were not shown in full to all participants.

\begin{figure*}[htbp]
    \centering
    \includegraphics[width=\textwidth]{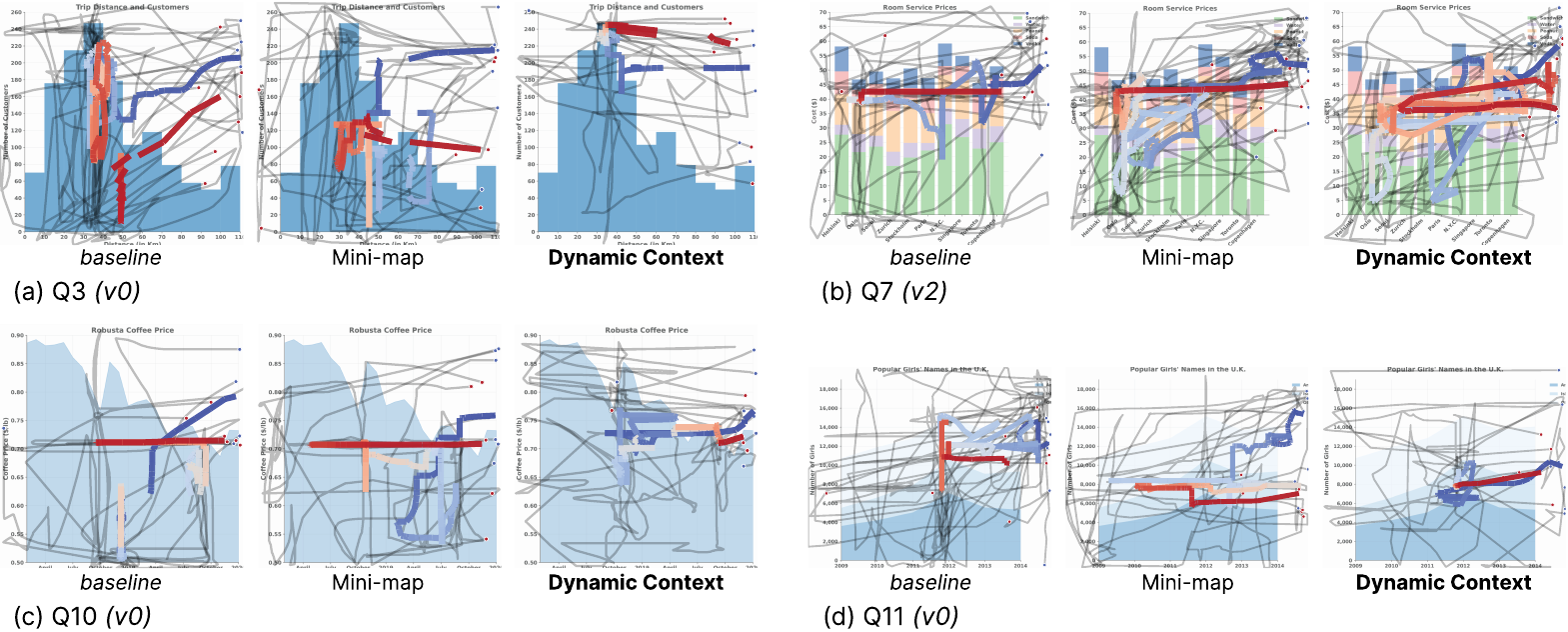}
        \caption{Comparison of significant trajectory differences between conditions. Individual trajectories are semi-transparent, with the mean trajectory highlighted.}
        \Description{Comparison of significant trajectory differences between conditions. Individual trajectories are semi-transparent, with the mean trajectory highlighted. Four charts are shown: Q3 (v0), Q7 (v2), Q10 (v0), and Q11 (v0). Each chart has three subfigures representing Baseline, Mini-map, and Dynamic Context conditions. The trajectories in the Dynamic Context condition are more centralized compared to the other two conditions.}
    \label{fig:trajectories_significant}
\end{figure*}

PERMANOVA revealed statistically significant differences in interaction patterns with \dc compared to \bl and \mm across several charts (\autoref{fig:trajectories_significant}). For \textbf{Q11 (v0; stacked area)}, \dc differed from \bl ($F=2.83$, $p=0.036$) and \mm ($F=3.47$, $p=0.006$); for \textbf{Q7 (v2; stacked bar)}, from \bl ($F=5.37$, $p=0.030$); for \textbf{Q3 (v0; histogram)}, from \mm ($F=4.47$, $p=0.030$); and for \textbf{Q10 (v0; area)}, from \mm ($F=3.27$, $p=0.048$).
In Q3 (\autoref{fig:trajectories_significant}.a), Q10 (\autoref{fig:trajectories_significant}.c), and Q11 (\autoref{fig:trajectories_significant}.d), \dc reduced panning actions, with Q11 showing fewer visits to the legend region. This aligns with observations that easier access to contextual information reduces reliance on panning, making chart reading a \q{more compact} experience (P8). P6 remarked, \q{I really, really, really liked the fact that you could bring the details near and then also have that crosshair to actually create a pinpoint for where that data is, that's amazing.}
Conversely, Q7 (\autoref{fig:trajectories_significant}.b) showed increased panning under \dc relative to \bl. As Q7 was among the most challenging charts (\autoref{tab:test_statistics}), the additional access provided by \dc encouraged more interaction and repeated solution attempts.

Participants used the crosshair as a visual anchor to track and compare data, and expressed that using the crosshair increased their confidence. P2 explained, \q{I really liked the crosshair. That was a really helpful guide when I was trying to determine the exact number of something. Because when I was working without any sort of guidelines or any assistance, I was using my mouse cursor to sort of, like, mimic that effect. I felt like if I had a fourth time to do it with the crosshairs, it would just be like, bam, bam, bam.}  
For some participants, the crosshair alongside the OA had a stabilizing effect (P9, P10, P12, P13). P9 expressed, \q{It helps me keep the focus, and so that way, I can kind of make sure my eyes aren't drifting, or filling in information that isn’t really there.}
A few participants (P12, P19) used the crosshair to compare and estimate scale, such as in the pie or stacked area charts. In the pie chart, participants placed the pointer at the center of the pie to compare the scale of slices with the four quadrants the crosshair created.

Since \dc uses a compact chart representation, some found the view to be \q{a little bit of clutter that made it feel un-intuitive and harder to read} (P1). P5 echoed this tension: \q{So the box kind of like helps in a sense that it brings the access to you, but because it's new and it makes more clutter on the screen, it can also be kind of challenging or difficult to get used to.}
For participants with double vision (\eg P6, P11, P21) keeping track was more challenging. P6 noted, \q{It did get a little overwhelming at times, especially with my double vision, but it was still worth it because I could then actually fall back} to their regular workflow.
Participants suggested activating \dc on demand (\eg P5, P8, P10), adding borders to projected elements (P6), toggling visibility of individual elements (P3, P6), dimming other content (P5), or even removing duplicated content from the chart (P1), as ways to address this challenge. As P4 summarized, \q{It’s tricky because it could be too many things in my view, too busy, and I would be lost. But it’s about balance, which is kind of personal.} 

\paragraph{\textbf{Using \mm (overview+detail).}}
Participants would use the minified view of charts to find the next point of interest or get global context for orientation. 
P4 commented, \q{I used it more to orient myself in the graph because of my magnification, but my magnification level is not so high.} Others found it useful in visually sparse areas; P5 explained, \q{If I'm zoomed into a place without any plotline, if I really didn't have any reference, I wouldn't know if the plotline is above or below.} 
Similarly, P18 explained, \q{In one or two cases, it was immediately very helpful in giving context. [about the bubble chart] I looked for what seemed to be the data point furthest to the left. Then I zoomed-in to read the name. And I can glance at the mini-map and see, yes, this is the leftmost point.}

Like \dc, some participants used the OA of \mm to track and compare values (\S\ref{sec:implementation}). For example, P6 and P9 widened and shortened the area and used it like a ruler. P9 noted, \q{I was using that border to align things.}
However, most participants found \mm offered limited benefit within their current strategies. Most screen magnifier users first zoomed out or physically stepped back to locate an area of interest, then zoomed in to inspect details; thus the orientation support \mm delivered often felt redundant. P3 explained, \q{Because I was zooming in on a known position, I didn't need to know where my dot was. I’ve been using zoom for a long time, and that’s my method, so I didn’t find any gain in the small map.} 

While \mm adapted overview+detail to LVI by using a compact layout and customization settings, some participants still found it less accessible (\eg P3, P6, P9, P14, P22).
P6 explained, \q{It's not for me because it's just so small [...] There was one time where I could tell that one [bar] was taller, but it's too small for it to be useful.}
Although participants acknowledged the potential benefits of \mm in other contexts, it was not effective for this task. P9 noted, \q{If I was at my desk in my office, I have a TV and I get lost on that screen,} as a potential scenario that \mm would be helpful in.
Therefore, \mm did not significantly contribute to or change the strategies of participants. 

\subsubsection{Usability and workload (RQ4)}
Our prototype interaction methods were more usable and reduced the overall effort. 
Participants rated \dc as significantly more usable for chart-reading tasks compared to the other conditions.  
SUS scores were highest for \dc ($M = 76.36$, $SD = 17.62$), compared to \mm ($M = 61.82$, $SD = 22.20$) and \bl ($M = 52.39$, $SD = 20.89$). Statistical analysis revealed a significant effect of Condition on SUS (LME: $F(2, 32) = 9.264$, $p < .001$, $\eta^2_p = 0.367$), with post-hoc comparisons showing that \dc was significantly higher than \bl ($p < .001$).
Overall, most reported they preferred using \dc; P11 and P17 preferred using \mm; however, P7 and P8 preferred using neither.

\begin{figure}[htbp]
    \centering
    \includegraphics[width=0.95\linewidth]{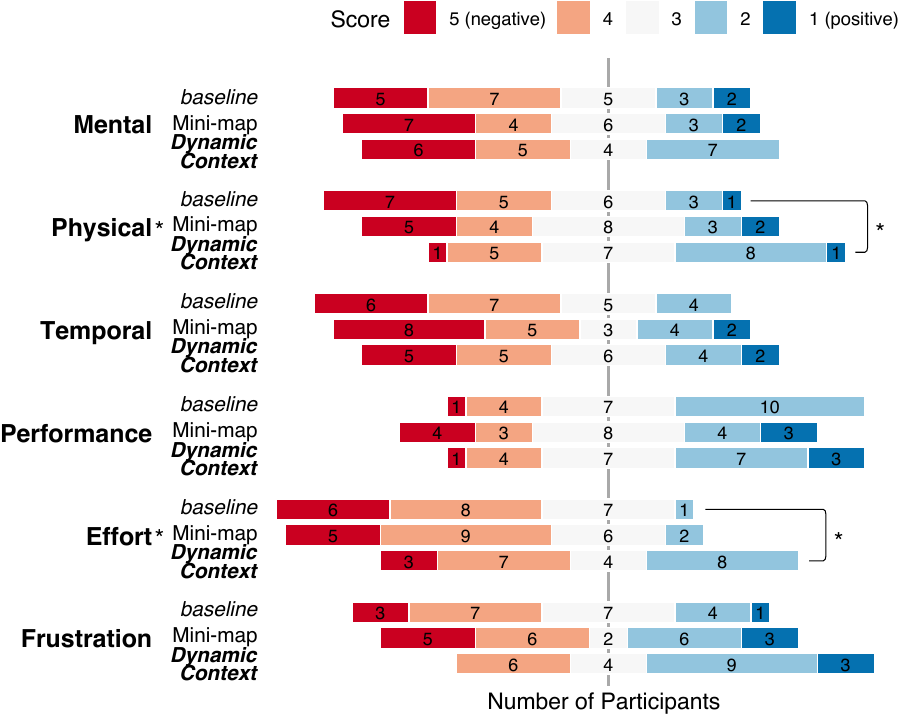}
    \caption{Task load evaluation results. Asterisks (\textit{*}) indicate statistically significant differences (p < 0.05).}
    \label{fig:task_load}
    \Description{Task load evaluation results. We see significant improvements in Effort and Physical Demand.}
\end{figure}

Task load evaluation reveals a significant improvements in \textit{Effort} and \textit{Physical Demand} (\autoref{fig:task_load}).  
Condition had a significant effect on Effort (LME: $F(2, 32) = 4.239$, $p = 0.023$, $\eta^2_p = 0.209$), with post-hoc comparisons indicating that \bl required more effort than \dc ($p = 0.027$). Similarly, Condition significantly affected Physical Demand (LME: $F(2, 32) = 3.606$, $p = 0.039$, $\eta^2_p = 0.184$), with \bl being more physically demanding than \dc ($p = 0.030$).
Qualitative feedback reflected these findings. Participants noted the excessive movements required with \bl. For example, P2 explained, \q{I moved a lot more than other times, because I had to just keep moving my head to try and keep track of things I see.} P7 expressed a similar sentiment, \q{It was kind of difficult. I have to go back and forth on the screen.}  

\begin{figure}[htbp]
    \centering
    \includegraphics[width=\linewidth]{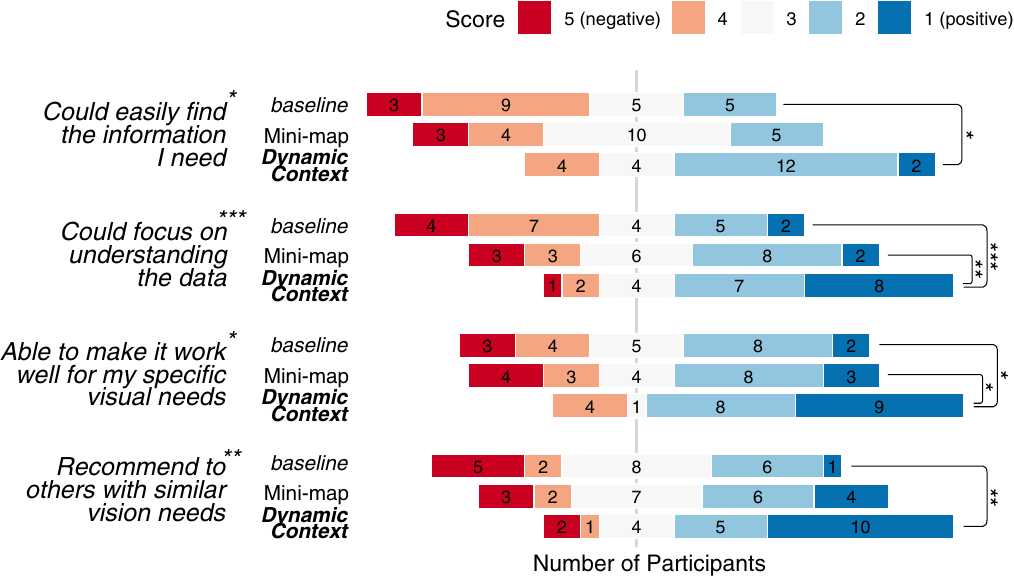}
    \caption{Experience evaluation results. Asterisks indicate statistical significance: \textit{*} $p<0.05$, \textit{**} $p<0.01$, \textit{***} $p<0.001$.}
    \label{fig:experience}
    \Description{Experience evaluation results. We see that Dynamic Context is rated significantly higher than both Baseline and Mini-map on all metrics, indicating a better overall user experience.}
\end{figure}

\subsubsection{Access to charts (RQ5)}
We evaluated four sense-of-access metrics across conditions. Participants rated \dc significantly higher than \bl on all dimensions, and higher than \mm on some (\autoref{fig:experience}).  
Participants found \dc most effective for engaging with their data. Condition had a significant effect on \textit{ability to focus on the data} (LME: $F(2, 32) = 7.505$, $p = 0.002$, $\eta^2_p = 0.319$), with \dc enabling better focus than \bl ($p = 0.002$). Similarly for \textit{sense of access to graph information} (LME: $F(2, 32) = 10.029$, $p < .001$, $\eta^2_p = 0.385$), where \dc outperformed both \bl ($p < .001$) and \mm ($p = 0.009$).
Participants also reported a stronger sense of compatibility with \dc. We found significant effects for both \textit{fitting with participants' needs} (LME: $F(2, 32) = 5.196$, $p = 0.011$, $\eta^2_p = 0.245$) and \textit{degree of recommendation} (LME: $F(2, 32) = 6.093$, $p = 0.006$, $\eta^2_p = 0.276$). For personal fit, \dc was preferred over both \bl ($p = 0.016$) and \mm ($p = 0.036$), while for recommendability, participants were significantly more likely to recommend \dc than \bl ($p = 0.004$). 

Participant feedback shows that \dc increased confidence and sense of agency.
The crosshair, in particular, was valued for accuracy in complex visual tasks. P3 explained, \q{The crosshairs gave me more control, thereby more confidence, thereby more likelihood of getting the answer correct if I had sufficient time.}
For some participants, using \dc provided a sense of visual agency.
P6 noted, \q{This tool forces you to actually do the work, it does not give you the answer. It gives you the answer in the sense of how a sighted person has the answer because they can see the lines without assistance.} Similarly, P2 reflected, \q{That was one of those like, 'oh, is this what sighted people get to experience when they do math?' [...] I feel that it gave me a glimpse of what sighted people experience, because the information was right there. I wasn't having to desperately hope I could find it.}  
P10 even suggested that \dc could benefit everyone, stating, \q{I think this kind of thing would help other people, even people who are not visually impaired.}

Despite the context provided by \dc and \mm, some context remained inaccessible, such as the scope of the available information itself. Users struggled to ascertain if an answer or specific data element was present but currently obscured, or genuinely absent from the chart (i.e., the ``not knowing what you don't know''~\cite{bigham2014MakingWeb}).
For instance, when using \dc, P5 commented, \q{On some questions, I want to know if the information is even available in the chart or not.} This difficulty in establishing visible scope also led to confusion, as P3 felt confused when the percentages on a pie chart were not visible.
Moreover, a few participants (\eg P7, P8, P11) expressed limited trust in both interaction prototypes. P8 explained, \q{I’m still trying to get a feel for it, what it's actually doing. I kind of don't trust it for some reason. I like to find answers myself.}

%% file: tables/05-participant-table.tex
\begin{table*}[t]
  \caption{Participant Demographics. \textbf{A/G}: Age and Gender. \textbf{CVD}: Color Vision Deficiency (Low Cont. = Low Contrast, Light Sens. = Light Sensitivity). \textbf{O}: Onset (A=Acquired, C=Congenital). \textbf{LB}: Recognized as Legally Blind. \textbf{MS}: Monitor Size. \textbf{CF}: Charts Familiarity. \textbf{Diagnoses}: ROP (Retinopathy of Prematurity), RP (Retinitis Pigmentosa), PDR (Proliferative Diabetic Retinopathy). \textbf{Gender}: M (Male), F (Female), NB (Non-binary), AG (Agender). Visual Acuity/Field indicate both left (L) and right (R) eyes when available. $\dagger$ indicates participation in formative study.}
  \label{tab:study_participants}
  \centering
  \begin{tabular}{l l l l l c l c c c}
    \toprule
    \textbf{PID} & \textbf{A/G} & \textbf{Diagnosed Condition} & \textbf{Visual Acuity} & \textbf{Visual Field} & \textbf{O} & \textbf{CVD} & \textbf{LB} & \textbf{MS}  & \textbf{CF}\\
    \midrule
    P1 & 49/NB & Bilateral optic neuropathy & R:20/400, L:20/250 & 10$^{\circ}$, central loss & A & Low Cont. & \textbf{T} & 27" & 5 \\
    P2 & 43/AG & ROP & R:None, L:20/500 & Peripheral loss & C & --- & \textbf{T} & 27" & 3 \\
    P3 & 66/M & Optic neuropathy & 20/1000 & Intact & C & Low Cont. & \textbf{T} & 27" & 3 \\
    P4 & 28/M & Aniridia, photophobia & R:20/120, L:20/200 & Intact & C & --- & \textbf{T} & 32" & 3 \\
    P5 & 35/M & Albinism, nystagmus, myopia & 20/200 & Intact & C & Light Sens. & \textbf{T} & 15" & 5 \\
    P6 & 39/F & PDR, double vision & R:20/50, L:20/200 & Peripheral loss & A & --- & F & 19" & 4 \\
    P7 & 41/M & ROP & R:No light, L:5/400 & 10$^{\circ}$ & C & Low Cont. & \textbf{T} & 48" & 5 \\
    P8 & 59/F & Optic/Solar neuropathy & 20/200 & Intact & A & --- & \textbf{T} & 24" & 5 \\
    P9 & 60/F & Glaucoma, astigmatism & 20/30 & R:5$^{\circ}$, L:finger motion & A & --- & \textbf{T} & 15" & 5 \\
    P10 & 29/M & RP, photophobia & Unknown & $<10^{\circ}$ & C & Low Cont. & \textbf{T} & 27" & 4 \\
    P11 & 29/M & Blue cone monochromatism & R:20/200, L:20/350 & Central loss & C & Yes & \textbf{T} & 16" & 4 \\
    P12 & 25/F & Albinism, nystagmus & 20/40 & Intact & C & --- & F & 27" & 4 \\
    P13 & 52/M & Pituitary tumor & Unknown & R:10$^{\circ}$, L:2$^{\circ}$ & A & --- & \textbf{T} & 22" & 5 \\
    P14 & 23/F & Brain tumor & Unknown & Limited (60$^{\circ}$) & A & --- & \textbf{T} & 31" & 4 \\
    P15 & 66/M & Cone-rod dystrophy & 20/600 & Limited (120$^{\circ}$) & A & Yes & \textbf{T} & 24" & 4 \\
    P16 & 30/M & RP & 20/100 & Limited (120$^{\circ}$) & C & Yes & F & 32" & 4 \\
    P17 & 63/M & Cataracts & R:20/400, L:No light & Intact & C & --- & F & 17" & 2 \\
    P18$\dagger$ & 32/M & Bilateral juv glaucoma, cataracts & R:20/200, L:No light & Peripheral loss & C & --- & \textbf{T} & 32" & 5 \\
    P19 & 28/M & Nystagmus & R:20/150, L:20/125 & Intact & C & Red-Green & \textbf{T} & 15" & 5 \\
    P20$\dagger$ & 28/F & Strabismus, Chiari malformation & R:20/250, L:20/320 & Limited (65$^{\circ}$) & C & Low Cont. & \textbf{T} & 23" & 5 \\
    P21 & 43/F & Retinal disease, cataracts & 20/400 & Peripheral loss & A & Low Cont. & \textbf{T} & 24" & 4 \\
    P22$\dagger$ & 42/F & Albinism, nystagmus, astigmatism & Unknown & Intact & C & Light Sens. & \textbf{T} & 27" & 5 \\
    \bottomrule
  \end{tabular}
\end{table*}

%% file: tables/05-test-statistics.tex
\begin{table*}[htbp]
\caption{Statistics of responses for each outcome type (\cmark=correct, \xmark=incorrect, TO=timeout, SK=skipped) across conditions (\bl, \mm, \dc) for all test questions Q1-Q12.}
\label{tab:test_statistics}
    \centering
\begin{tabular}{llcccc|cccc|cccc}
    \toprule
    \multirow{2}{*}{\textbf{QID}} & \multirow{2}{*}{\textbf{Chart}} & \multicolumn{4}{c}{\textbf{\bl}} & \multicolumn{4}{c}{\textbf{\mm}} & \multicolumn{4}{c}{\textbf{\dc}} \\
    \cmidrule(lr){3-6} \cmidrule(lr){7-10} \cmidrule(lr){11-14}
     & & \cmark & \xmark & \textbf{TO} & \textbf{SK} & 
       \cmark &  \xmark & \textbf{TO} & \textbf{SK} & 
       \cmark & \xmark & \textbf{TO} & \textbf{SK} \\
    \midrule
    \textbf{Q1} & Treemap & 100.0\% & 0\% & 0\% & 0\% & 81.8\% & 9.1\% & 9.1\% & 0\% & 95.5\% & 0\% & 4.5\% & 0\% \\
    \textbf{Q2} & 100\% stacked bar & 95.5\% & 4.5\% & 0\% & 0\% & 86.4\% & 4.5\% & 4.5\% & 4.5\% & 77.3\% & 18.2\% & 4.5\% & 0\% \\
    \textbf{Q3} & Histogram & 86.4\% & 9.1\% & 4.5\% & 0\% & 59.1\% & 27.3\% & 13.6\% & 0\% & 77.3\% & 9.1\% & 13.6\% & 0\% \\
    \textbf{Q4} & Choropleth & 72.7\% & 4.5\% & 13.6\% & 9.1\% & 59.1\% & 9.1\% & 22.7\% & 9.1\% & 59.1\% & 9.1\% & 13.6\% & 18.2\% \\
    \textbf{Q5} & Pie & 68.2\% & 31.8\% & 0\% & 0\% & 63.6\% & 31.8\% & 4.5\% & 0\% & 54.5\% & 40.9\% & 4.5\% & 0\% \\
    \textbf{Q6} & Bubble & 63.6\% & 22.7\% & 4.5\% & 9.1\% & 50.0\% & 31.8\% & 13.6\% & 4.5\% & 45.5\% & 40.9\% & 13.6\% & 0\% \\
    \textbf{Q7} & Stacked bar & 4.5\% & 45.5\% & 22.7\% & 27.3\% & 18.2\% & 27.3\% & 40.9\% & 13.6\% & 4.5\% & 36.4\% & 45.5\% & 13.6\% \\
    \textbf{Q8} & Line & 77.3\% & 18.2\% & 4.5\% & 0\% & 72.7\% & 18.2\% & 9.1\% & 0\% & 68.2\% & 22.7\% & 9.1\% & 0\% \\
    \textbf{Q9} & Bar & 95.5\% & 4.5\% & 0\% & 0\% & 81.8\% & 13.6\% & 4.5\% & 0\% & 95.5\% & 4.5\% & 0\% & 0\% \\
    \textbf{Q10} & Area & 63.6\% & 27.3\% & 9.1\% & 0\% & 72.7\% & 22.7\% & 4.5\% & 0\% & 54.5\% & 36.4\% & 9.1\% & 0\% \\
    \textbf{Q11} & Stacked area & 27.3\% & 22.7\% & 27.3\% & 22.7\% & 27.3\% & 36.4\% & 22.7\% & 13.6\% & 13.6\% & 40.9\% & 31.8\% & 13.6\% \\
    \textbf{Q12} & Scatter & 54.5\% & 18.2\% & 9.1\% & 18.2\% & 59.1\% & 9.1\% & 13.6\% & 18.2\% & 68.2\% & 13.6\% & 9.1\% & 9.1\% \\
    \bottomrule
\end{tabular}
\end{table*}

%% file: content/06-discussion.tex
\section{Discussion}
\label{sec:discussion}

In the previous section (\S\ref{sec:user-study}), we evaluated how our proposed interaction methods performed on a chart question-answering task using 12 types of charts. 
In this section, we reflect on our results (\S\ref{sec:results}) and categorize them into three themes, which we present as design insights (DIs) for future systems that provide visual context for LVI.
Following that, we discuss how our approach can extend to include higher-level context---such as trends---and be adapted to other domains beyond charts.
Finally, we reflect on our study and discuss limitations and avenues for future work.

\subsection{Design Insights to Deliver Context for LVI}
\dc delivered a significant benefit for chart reading (\S\ref{sec:results}). 
Although completion times and scores did not reflect this, subjective evaluations, qualitative feedback, and our observations supported this finding.
Conversely, the overview+detail approach of \mm, while outperforming the \bl, was less useful than \dc and provided less access in this task---corroborating prior works~\cite{hornbaek2002NavigationPatterns}.
Solving the Mini-VLAT with \dc required significantly less effort and led to a significantly higher sense of access. 
For instance, participants interacted more with the chart data and navigated to the axes or legend less.
However, the current prototype implementation of \dc and \mm was not always compatible with current AT use, the design of \dc increased participants' visual overload, and many participants found \mm too small and inaccessible.
Despite this, our results indicate a positive effect for systems that provide visual context for LVI.

From our findings, we surface three emerging themes that we present as DIs to guide the development of future systems for delivering visual context access to LVI.
While our insights are motivated by the needs of LVI, improving visualization accessibility for them can, in turn, benefit everyone~\cite{wimer2024VisionImpairments,marriott2021InclusiveData}.

\subsubsection{Providing visual context increases visual access (DI1)}
Both of our proposed interaction prototypes use visuals to convey context---for example, projecting the axes around the pointer.
We enable access to spatial context for tracking and comparing data (e.g., grid lines) using visual anchors in the form of a crosshair.
The crosshair functionality provided more control when tracking or comparing data values. For instance, participants used it to align the crosshair with the center of a pie chart, the tops of bars, or to estimate the exact value of a point on a line chart. 
The legend was consistently placed at the top-right, and beyond the context of data categories, participants also used it to compare colors in complex situations such as continuous scales (e.g., choropleth); however, it did not assist with other forms of CVD.
For LVI, visual access (more than non-visual access) is inherent to an accessible experience~\cite{szpiro2016HowPeople}. With accessible visual context LVI could emulate the \q{sighted experience} (P2).
Current AT does not provide workflows comparable to what sighted people experience~\cite{zong2022RichScreen}; however, both LVI and sighted people approach charts via similar underlying visual instincts~\cite{shneiderman1996EyesHave}.
Adding \dc to existing AT introduced access to visual context, and to additional workflows for reading charts.

\subsubsection{Reducing visual complexity is necessary for visual access (DI2)}
One core challenge with \dc and \mm was visual complexity. In \dc, we superimpose context into a compact area around the pointer. This resulted in an interface with duplicated chart elements and high information density, which was especially harmful for participants with double vision (\eg P6, P20), or who use high levels of magnification such as 400\% or above (\eg P3). 
Limiting what context is displayed can improve this. For instance, strategies include adding borders to elements~\cite{elavsky2022HowAccessible}, toggling the visibility of individual elements, or even removing the original content from the chart.
\mm displays a smaller copy of the chart; however, it was not accessible. 
For an effective visual aid to convey spatial context, \mm should include fewer details.
In their seminal work, \citet{treisman1980FeatureintegrationTheory} show how the most instantaneous perception we have of visuals are visual \textit{features} (\eg color, shape, texture). To understand what we see, we combine these features into \textit{elements} which contain meaning. 
For LVI, limitations of visual perception reduce the scope of noticeable features. For instance, LVI find it nearly impossible to notice pointers in presentation videos~\cite{sechayk2025VeasyGuidePersonalized}.
Therefore, spatial context via an overview~\cite{hornbaek2002NavigationPatterns} should generate an abstracted representation of the chart.
For example, on a bar chart, this could mean removing text elements, using distinct spacing between bars, and indicating the rough position of the title and the legend.

\subsubsection{Support variability of fluctuating user needs (DI3)}
User needs are not limited to their visual condition, but include: \q{What is their current goal?}, \q{What content are they viewing now?}, and \q{What AT configuration are they using?} 
For instance, users may have different information goals---such as finding the highest value or interpreting trends in the data, change magnification levels, or view different charts. The context or its presentation should adapt to these changing needs.
In our study, \dc had no noticeable benefit on the scatterplot whose question asked about a relationship between the two axes. In some cases, the visual representation of \dc confused participants, who thought this detailed information was important to the question. Conversely, using the \mm participants could have access to the axes, with the overview accessible in their view.
Participants used various AT (\A\ref{apx:study-at-use}), and modified these during the study, influenced by in-the-moment needs~\cite{newell1995ExtraordinaryHumancomputer}. For example, they would increase the zoom level to read axis values, decrease the zoom level to find the next point of interest, or change display colors by toggling color inversion. 
However, our prototypes did not consider these changes. When P5 toggled color inversion, the outer dimming turned white, and when P3 changed magnification levels, the context in the OA was clipped.
In their work, \citet{wobbrock2011AbilitybasedDesign} presents \textit{ability-based design}. Systems based on ability-based design are intrinsically aware of user abilities, optimizing the interface for the user's functional profile. Adaptable systems shift the burden of adjustments from the user to the software. 
For instance, this involves changing the OA dimensions based on magnification levels, changing element colors based on system-wide color changes, or populating different default configurations based on which AT are in use. 
While personalization is a core implementation goal (consistent with other notable LVI works~\citep[\eg][]{zhao2015ForeSeeCustomizable,sechayk2025VeasyGuidePersonalized,sackl2020EnsuringAccessibility}), automatically adapting settings to user's situational needs is a natural next step towards making context accessible for LVI.

\subsection{High-level Context and Other Domains}
In this work, we provide access to low-level visual context (\eg, axes, legend, grid lines) and spatial context via the overview. However, we do not address higher-level context, such as trends or relationships in the data. 
Prior work addressed this high-level context using AI conversational agents~\cite{kim2023ExploringChart, gorniak2024VizAbilityEnhancing,seo2024MAIDRMeets}, providing access through audio descriptions or conversations.
Nevertheless, these approaches pose trust concerns among users due to the risk of AI hallucination~\cite{wolfel2024KnowledgebasedGenerativeAIdriven}. Furthermore, these solutions are designed primarily for non-visual access, contrasting with our results that show the importance of supporting LVI visual agency and navigation.
Future work could use visual augmentations of low-level chart context to ground AI-generated high-level context~\cite{lang2021PressingButton,sechayk2025VeasyGuidePersonalized}. For instance, when an increasing trend is detected in a bar chart, the tops of the bars that belong to this trend can be highlighted with a distinct color and connected via a series of arrows.

While our objective was to improve chart accessibility for LVI, our proposed interaction methods can be adapted to other domains, such as day-to-day computer use. 
For instance, while prior works have explored \mm-akin approaches for web pages or diagrams~\cite{burigat2008MapDiagram}, \dc could similarly enhance web page navigation by incorporating information on page structure and providing visual context, building upon existing methods for visual web navigation for LVI~\citep[\eg][]{billah2018SteeringWheelLocalitypreserving}.
Furthermore, \mm could provide spatial context beyond the current position. 
Similar to research with screen reader users~\cite{zong2022RichScreen,blanco2022OlliExtensible}, spatial context concerns hierarchical information, as well as the presence or absence of visual elements in and around the current view.
For instance, successful accessibility goes beyond merely informing LVI about what is present; it must also convey information regarding the absence of expected elements or context~\cite{bigham2014MakingWeb}.

\subsection{Limitations and Future Work}
Our study is subject to several limitations. First, we observed a learning effect influenced by participants' familiarity with the test charts' structure. This effect strengthened over the course of the study due to the similarity in chart structure between Mini-VLAT variants (\A\ref{apx:vlat-variants}). This reduced both the response time and the amount of pointer interaction they performed. We used counterbalancing to mitigate this; however, using completely different charts between conditions, or a between-subject setup, can be considered for further studies.
Additionally, with 22 participants using vastly different workflows for solving the tasks, our quantitative results are mixed. We prioritized external validity and, therefore, did not limit the tools or type of access participants chose. Future work could recruit a larger sample size and employ stratified analyses to better disentangle the effects of these diverse workflows and visual capabilities; however, this is a challenging task~\cite{mack2021WhatWe}.

We did not conduct a direct comparison with prior work~\cite{prakash2025EnhancingLow,alcaraz-martinez2024EnhancingStatistical}. This decision was made to minimize adverse effects on LVI participants, such as fatigue and eye strain, which are critical risks in studies involving this population~\cite{sechayk2025VeasyGuidePersonalized}. Furthermore, such a comparison is infeasible due to fundamental differences in scope; existing methods are either not designed for complex chart types like stacked area charts~\cite{prakash2025EnhancingLow} or do not center on chart context~\cite{alcaraz-martinez2024EnhancingStatistical}. 
Finally, a meaningful comparison is complicated by the cognitive load required for users to \q{unlearn} established workflows, which inevitably impacts initial interactions with new tools. 
During the study, several participants defaulted to their original strategy, later realizing the access our prototypes provide.
Assessing the long-term efficacy of \dc and \mm requires a longitudinal study to observe any sustained impact of usage, particularly the stabilization of performance and any resulting shifts in user strategy. Such a study should contrast their use with other implemented solutions.

%% file: content/07-conclusion.tex
\section{Conclusion}
Charts remain prevalent for presenting information, but reading them is a challenging visual task with multiple layers of complexity. Existing accessible approaches primarily rely on non-visual modalities (e.g., audio), but they often restrict the user's visual agency to explore data independently, even though LVI prefer using their vision. In a formative study with five LVI participants, we identified four contextual elements whose access is crucial for chart reading: axes, legend, grid lines, and the overview. To investigate how to best make these elements accessible, we compared two interaction methods: \dc and \mm. Both deliver context in a compact layout that follows the pointer and is designed to accompany existing AT. In an evaluation with N=22 LVI, \dc significantly reduced physical effort and provided greater perceived access to charts compared to \bl. Conversely, \mm aided orientation by providing spatial context. However, the visual elements in \dc created overload, and the visuals in \mm were too small to perceive. We distill our findings into design insights that advocate for (1) maximizing visual access to context, (2) reducing complexity through abstraction, and (3) adapting to the dynamic needs of LVI. These ideas may also apply to other complex visuals, like diagrams or web pages. Overall, contextual overlays can improve access while preserving users’ control, pointing toward tools that support rather than replace visual interaction.

\section*{Open Science}
All code and data is openly available. Visit the project website at \url{https://visual-context.github.io/} for more information. 

%% file: content/11-acknowledgment.tex
We are grateful to the anonymous reviewers for their insightful feedback, and to all participants whose contributions enriched this study.
We thank Chu Li, Lozus Zhang, and Lucy Jiang for their thoughtful comments and support.
Yotam Sechayk was supported by the Ministry of Education, Culture, Sports, Science and Technology (MEXT) Monbukagakusho Scholarship.
Mark Colley was supported by a Canon Research Fellowship.
This work was supported by the Institute for AI and Beyond of
the University of Tokyo. 
This work was partially supported by Japan Science and Technology Agency (JST) as part of Adopting Sustainable Partnerships for Innovative Research Ecosystem (ASPIRE), Grant Number JPMJAP2401.

%% file: content/99-appendix.tex
\section{User Study Details}
\label{apx:user-study}
We provide additional information on the procedure and results of the user study.

\subsection{Counterbalancing Table}
\label{apx:counterbalance}
\input{tables/05-counterbalance-study}

\subsection{Initial Interview Questions}
\label{apx:initial-questions}
Below are the questions used during the initial interview of the user study (\S\ref{sec:user-study}). We exclude demographic questions from this list (e.g., age or gender identity).
\begin{enumerate}
    \item What is your vision condition? Is it congenital or acquired?
    \item Could you share with us your degree of visual acuity and field of vision? (Optional)
    \item Are you recognized as legally blind?
    \item What assistive tools do you use when using your computer?
    \item How familiar are you with charts? From \textsc{one} very unfamiliar to \textsc{five} very familiar.
    \item What device are you using now, and What is the size of its display?
\end{enumerate}

\subsection{Post-study Interview Questions}
\label{apx:post-study-questions}
Below are the questions used to guide the semi-structured interview at the end of the user study (\S\ref{sec:user-study}). For convenience, we refer to \dc and \mm as ``tools'' when asking participants. 
\begin{enumerate}
    \item What is your impression with the tools you used today?
    \item Are there any specific challenges when using the tools?
    \item Are there graphs one tool was better than the other?
    \item Which tool would you prefer to use overall?
    \item Do you have any additional comments?
\end{enumerate}

\newpage

\subsection{Participants AT Use}
\label{apx:study-at-use}
\input{tables/05-study-AT-use}

\clearpage
\onecolumn

\subsection{Examples Test Charts and Questions}
\label{apx:example-charts}
\input{tables/05-example-visualizations}

\section{Original VLAT Material and Variants}
\label{apx:vlat-variants}
For our user study procedure we used the Visualization Literacy Assessment Test (VLAT)~\cite{lee2017VLATDevelopment,pandey2023MiniVLATShort}, an established test involving interacting with, and reading charts. 
The test we used is based on Mini-VLAT~\cite{pandey2023MiniVLATShort}, which is a minimalist iteration of VLAT. 
While the original VLAT contains 12 visualizations with multiple questions each, Mini-VLAT has one question per visualization. Prior work demonstrates that, while shorter, Mini-VLAT remains effective at assessing visualization literacy~\cite{pandey2023MiniVLATShort}. 
Therefore, to avoid eye-strain and fatigue, which are a common occurrence for LVI, we choose Mini-VLAT for our study.

To effectively evaluate \dc and \mm we used 3 variations of the Mini-VLAT question set; the original set reproduced by us to the best of our abilities, and two additional variants. 
Prior work, used variations of VLAT to address cultural or language barriers, for example, \citet{omelchenko2025CrosslinguisticCultural} adapted the test to different cultures, comparing direct translations and question set variations. 
In this work, we generated variants in order to prevent response memorization and encourage participants to interact with each new graph. 
Each variant was generated while prioritizing preservation of the semantics of the question and visualization set. While we acknowledge that variations of the test may interfere with correct evaluation of visualization literacy, the goal of this study is not to correctly assess literacy. Table~\ref{tab:vlat-variations} shows an example of original vs. alternative variation from the study visualization and question pairs.

\subsection{Variant Generation}
Beyond the original Mini-VLAT question set, we generated two additional variants, prioritizing semantics preservation. In our variant generation process we used a set of five modification techniques; three modify the underlying data, and two modify the question.

\subsubsection{Data Modifications} We included three types of data modifications. These changed the numeric data associated with each visualization.
\begin{itemize}
    \item \textbf{Permutation:} Shifting the order of data points.
    \item \textbf{Noise:} Adding random noise to data points.
    \item \textbf{Magnitude:} Multiplying the data points by some constant value.
\end{itemize}

\subsubsection{Question Modifications} 
We used two types of question modifications. These changed the topic or orientation of the questions.
\begin{itemize}
    \item \textbf{Topic:} Changing objects or subjects mentioned in the question. For instance, changing \emph{``Gold medals''} to \emph{``Bronze medals''}.
    \item \textbf{Orientation:} Changing the orientation of the question target. For example, changing \emph{``What is the largest''} to \emph{``What is the smallest''}.
\end{itemize}

\input{tables/xx-vlat-variant-sample}

%% file: tables/05-counterbalance-study.tex
\begin{table}[ht]
\caption{Experimental condition counterbalancing. Each row represents the task sequence for one order index. 
Test variants (v0-v2) were distributed such that each appears exactly twice for each condition, and all test variants are used for each order sequence.}
\label{tab:counterbalancing}
\centering
\small
\setlength{\tabcolsep}{3pt} 
\begin{tabular}{@{}lll@{}}
\toprule
\textbf{Task 1} & \textbf{Task 2} & \textbf{Task 3} \\
\midrule
\texttt{\bl} (v0) & \texttt{\dc} (v1) & \texttt{\mm} (v2) \\
\addlinespace[2pt]
\texttt{\dc} (v0) & \texttt{\mm} (v1) & \texttt{\bl} (v2) \\
\addlinespace[2pt]
\texttt{\mm} (v0) & \texttt{\bl} (v1) & \texttt{\dc} (v2) \\
\addlinespace[2pt]
\texttt{\mm} (v2) & \texttt{\dc} (v1) & \texttt{\bl} (v0) \\
\addlinespace[2pt]
\texttt{\dc} (v2) & \texttt{\bl} (v1) & \texttt{\mm} (v0) \\
\addlinespace[2pt]
\texttt{\bl} (v2) & \texttt{\mm} (v1) & \texttt{\dc} (v0) \\
\bottomrule
\end{tabular}
\end{table}

%% file: tables/05-study-AT-use.tex
\begin{table}[ht]
\caption{AT usage reported by study participants. The legend indicates AT used during study tasks (\textbullet) versus general usage not employed during the study ($\circ$). Definitions are as follows: \textbf{Content Zoom} covers browser-based zoom, pinch-to-zoom, or text under pointer enlargement; \textbf{Color Filters} covers full-screen filters (e.g., inversion); \textbf{Color Themes} covers high-contrast themes or dark mode extensions; and \textbf{Custom Pointer} covers varied colors, sizes, or effects. Excluded from the table are P3's specific use of \textit{read text under pointer} for Q\&A and P20's general reliance on alt-text, which was unavailable for study charts.}
\label{tab:study-at-use}
\centering
\small
\setlength{\tabcolsep}{4pt} 
\begin{tabular}{lR{1.2cm}R{1.1cm}R{1.1cm}R{1.1cm}R{1.1cm}}
\toprule
PID & Screen Magnifier & Content Zoom & Color Filters & Color Themes & Custom Pointer \\
\midrule
P1 & $\circ$ & $\bullet$ & $\bullet$ &  &  \\
P2 &  & $\bullet$ &  & $\bullet$ &  \\
P3 & $\bullet$ &  &  &  &  \\
P4 & $\bullet$ &  &  & $\circ$ &  \\
P5 & $\bullet$ &  & $\bullet$ & $\circ$ &  \\
P6 & $\bullet$ & $\bullet$ &  & $\circ$ &  \\
P7 & $\bullet$ & $\bullet$ &  &  & $\bullet$ \\
P8 & $\bullet$ & $\circ$ & $\circ$ &  &  \\
P9 &  & $\bullet$  &  &  &  \\
P10 &  & $\circ$ & $\bullet$ &  &  \\
P11 & $\bullet$ & $\circ$ & $\bullet$ & $\circ$ &  \\
P12 &  & $\bullet$ & $\circ$ &  &  \\
P13 &  & $\bullet$ &  & $\circ$ &  \\
P14 & $\bullet$ & $\bullet$ &  & $\circ$ & $\bullet$ \\
P15 & $\bullet$ &  &  &  &  \\
P16 & $\bullet$ &  &  &  &  \\
P17 & $\bullet$ &  &  &  &  \\
P18 & $\bullet$ &  &  & $\bullet$ &  \\
P19 &  & $\bullet$ &  & $\circ$ & $\bullet$ \\
P20 & $\bullet$ & $\bullet$ & $\circ$ & $\circ$ & $\bullet$ \\
P21 &  & $\bullet$ & $\bullet$ & $\circ$ &  \\
P22 & $\bullet$ & $\bullet$ & $\circ$ & $\bullet$ & $\bullet$ \\
\bottomrule
\end{tabular}
\end{table}

%% file: tables/05-example-visualizations.tex
\begin{table*}[htbp]
    \caption{Questions used in the example quiz, which participants used to get accustomed with the tools and interface. Each question is accompanied by its answer options and graph. The correct answer is \colorbox{green!20}{highlighted}.}
    \label{tab:study_example_questions}
    \centering
    \small
    \begin{tabular}{p{0.30\textwidth}p{0.33\textwidth}p{0.30\textwidth}}
        \toprule
        \textbf{Question 1} & \textbf{Question 2} & \textbf{Question 3} \\ 
        \midrule
        \textbf{Q:} There are less cars with manual transmissions. & 
        \textbf{Q:} How many cars have 5 gears? & 
        \textbf{Q:} Which car weight (in 1000 lbs) has the lowest Miles per Gallon? \\ 
        \textbf{A:} \colorbox{green!20}{True}, False & 
        \textbf{A:} \colorbox{green!20}{5}, 8, 10, 12 & 
        \textbf{A:} \colorbox{green!20}{2.77}, 2.2, 2.72, 2.52 \\ 
        \vspace{0pt}\includegraphics[width=0.28\textwidth]{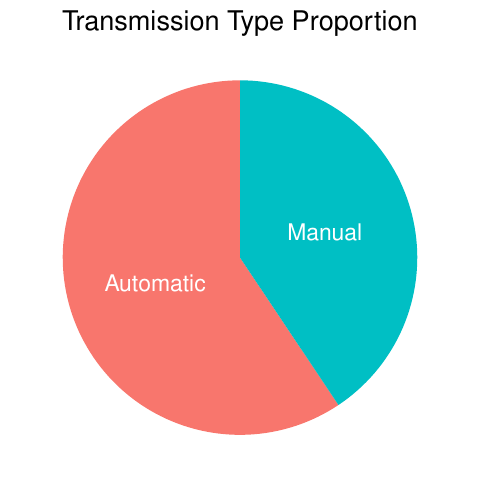} & 
        \vspace{0pt}\includegraphics[width=0.32\textwidth]{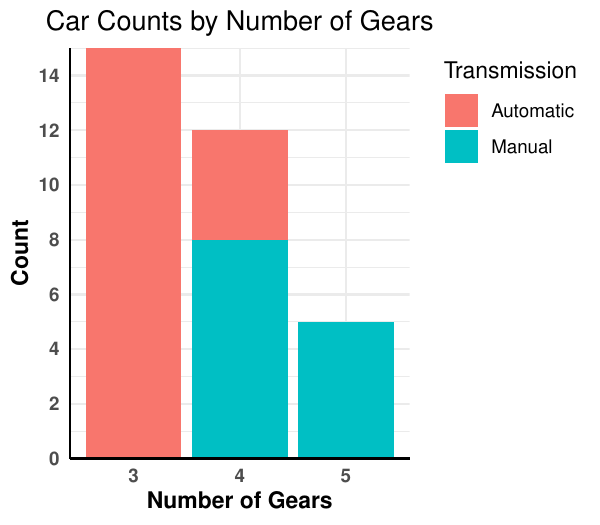} & 
        \vspace{0pt}\includegraphics[width=0.28\textwidth]{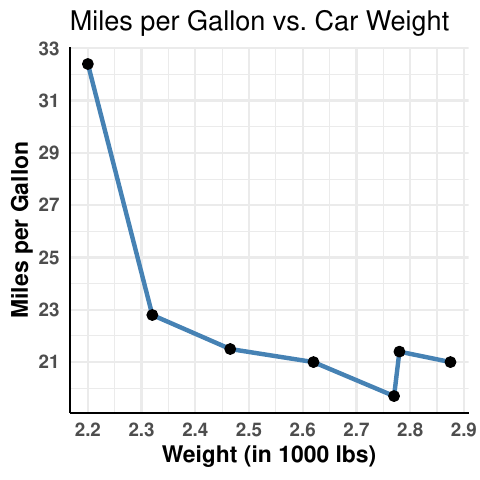} \\
    \bottomrule
    \end{tabular}
\end{table*}

%% file: tables/xx-vlat-variant-sample.tex
\begin{table*}[htbp]
    \caption{Example questions from the quiz variants used in the study. Each example shows the question, the graph, and the answers options. The correct answer is \colorbox{green!20}{highlighted}. For a comprehensive comparison of all questions see the supplementary material.}
    \label{tab:vlat-variations}
    
    \begin{tabular}{p{0.31\textwidth}p{0.31\textwidth}p{0.31\textwidth}}
        \toprule
        \textbf{Original} (v0) & 
        \textbf{Variant 1} (v1) & 
        \textbf{Variant 2} (v2) \\ 
        \midrule
        \textbf{Q:} Which country has the lowest proportion of Gold medals? & 
        \textbf{Q:} Which country has the lowest proportion of Bronze medals? & 
        \textbf{Q:} Which country has the highest proportion of Silver medals? \\ 
        \textbf{A:} USA, \colorbox{green!20}{Great Britain}, Japan, Australia & 
        \textbf{A:} Germany, Canada, \colorbox{green!20}{South Korea}, Brazil & 
        \textbf{A:} Germany, Great Britain, Japan, \colorbox{green!20}{Indonesia} \\ 
        \includegraphics[width=0.28\textwidth]{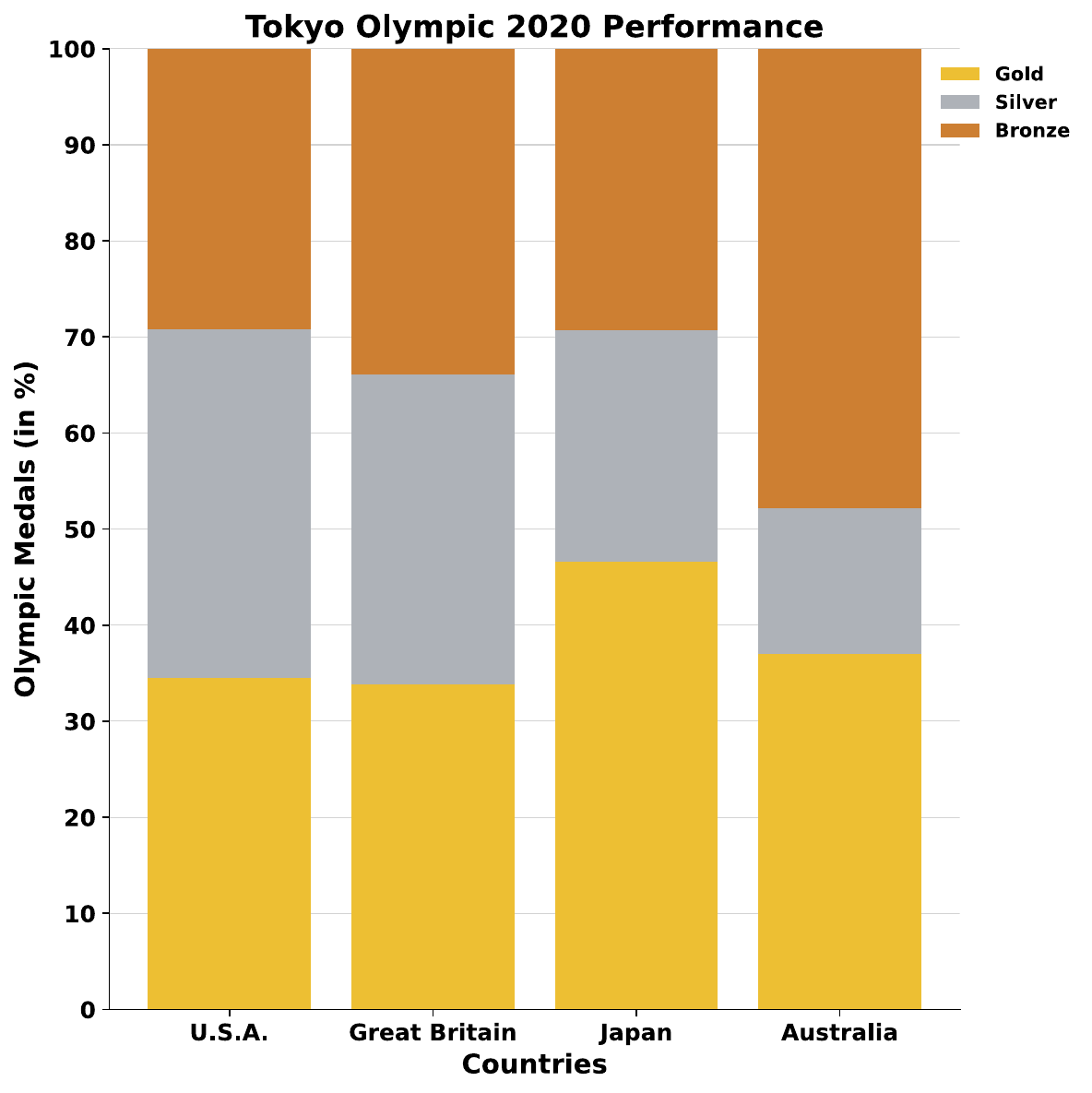} & 
        \includegraphics[width=0.28\textwidth]{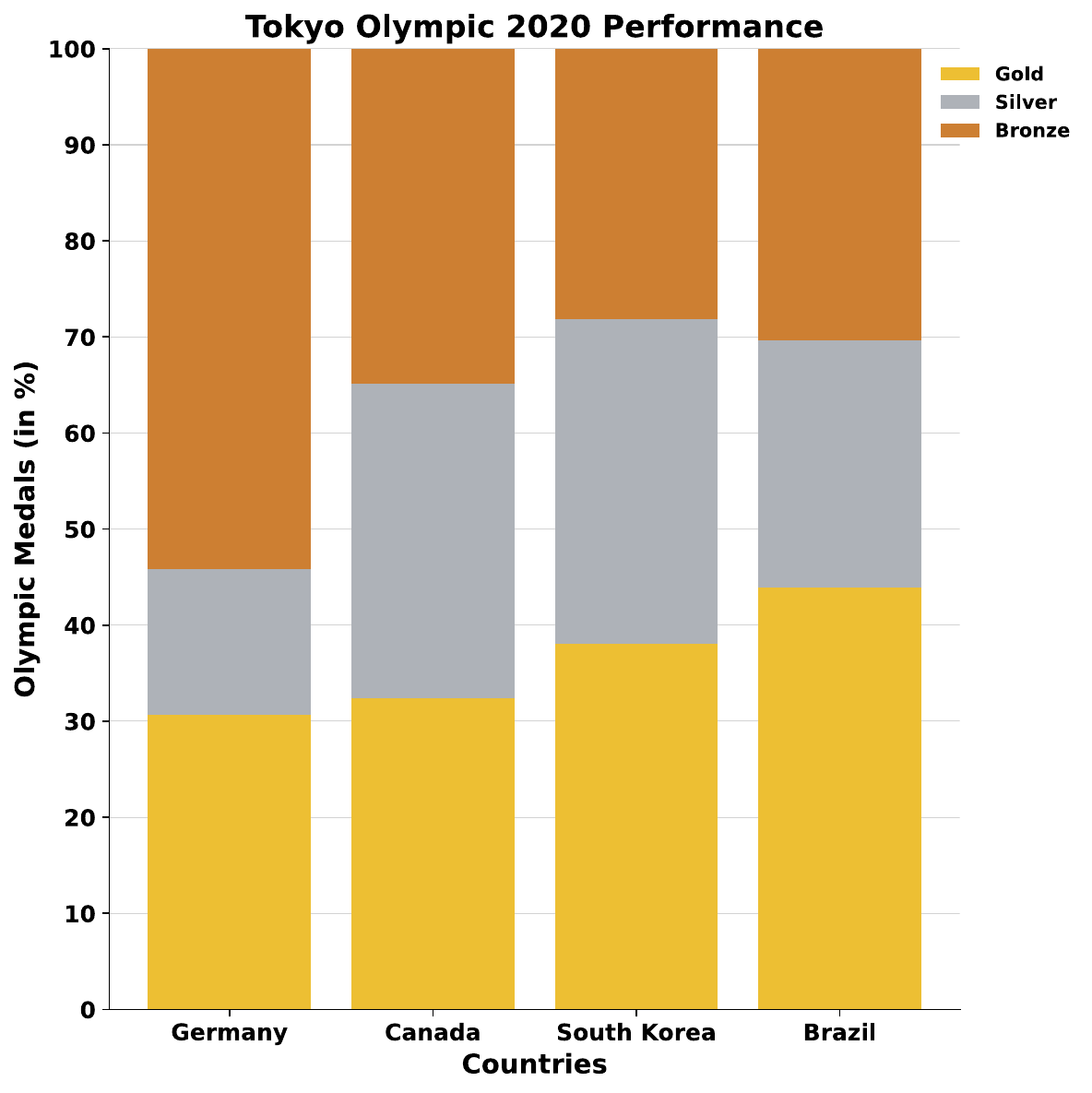} & 
        \includegraphics[width=0.28\textwidth]{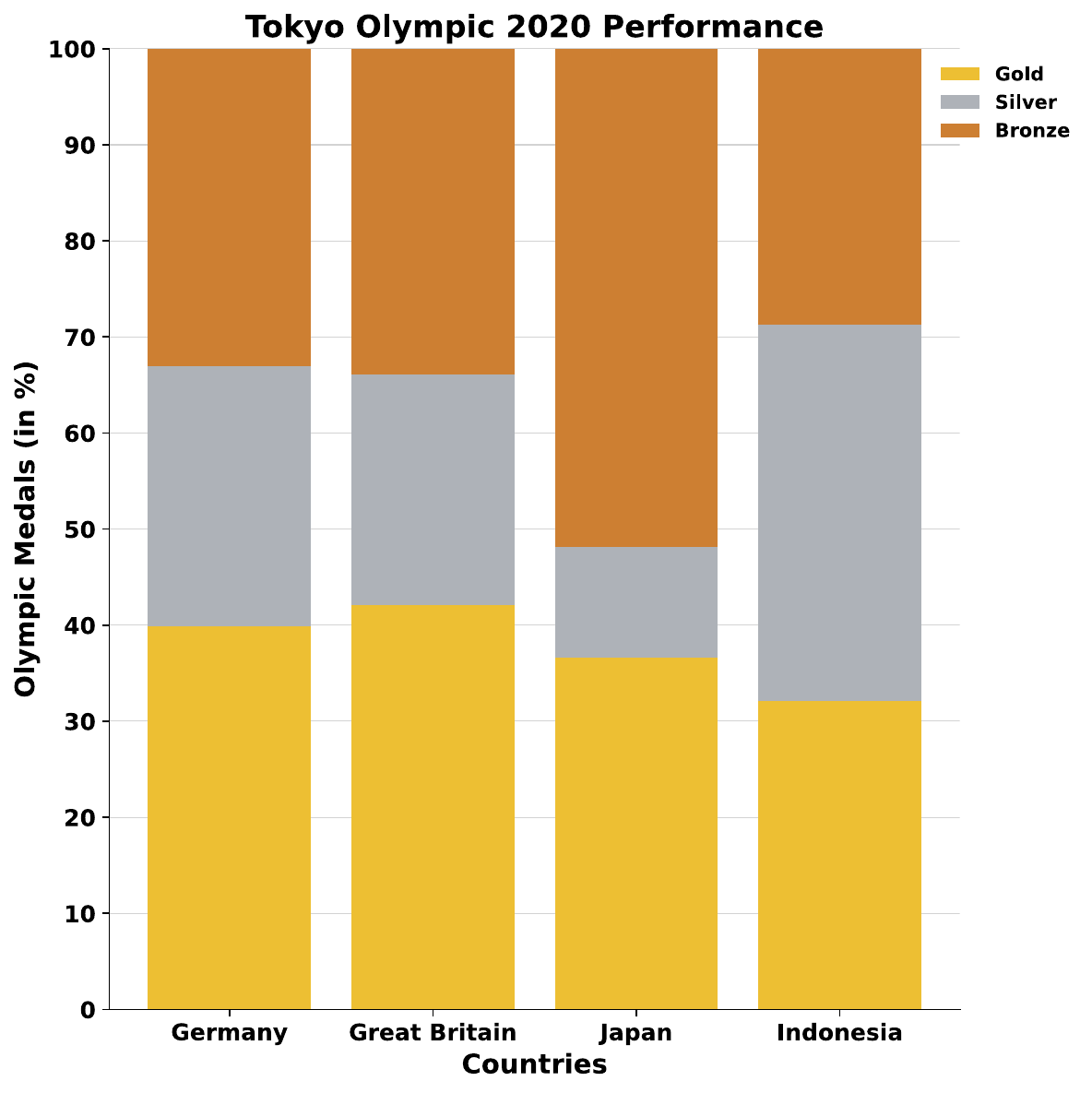} \\
        \midrule
        \textbf{Q:} What distance have customers traveled the most? & 
        \textbf{Q:} What distance have customers traveled the most? & 
        \textbf{Q:} What distance have customers traveled the least? \\ 
        \textbf{A:} 60–70 km, \colorbox{green!20}{30–40 km}, 20–30 km, 50–60 km & 
        \textbf{A:} \colorbox{green!20}{60-70 km}, 30-40 km, 20-30 km, 50-60 km & 
        \textbf{A:} 60-70 km, 30-40 km, \colorbox{green!20}{100-110 km}, 50-60 km \\ 
        \includegraphics[width=0.28\textwidth]{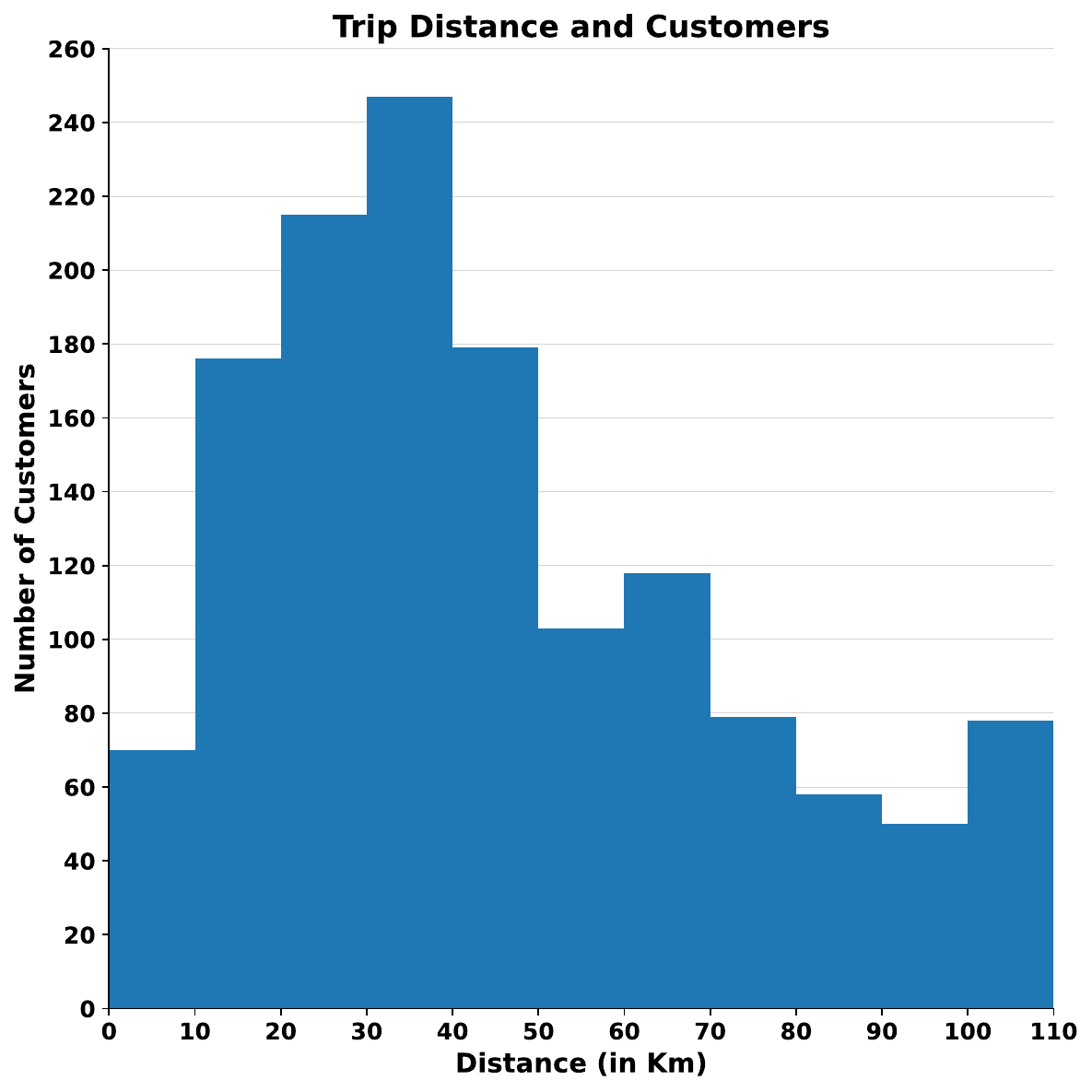} & 
        \includegraphics[width=0.28\textwidth]{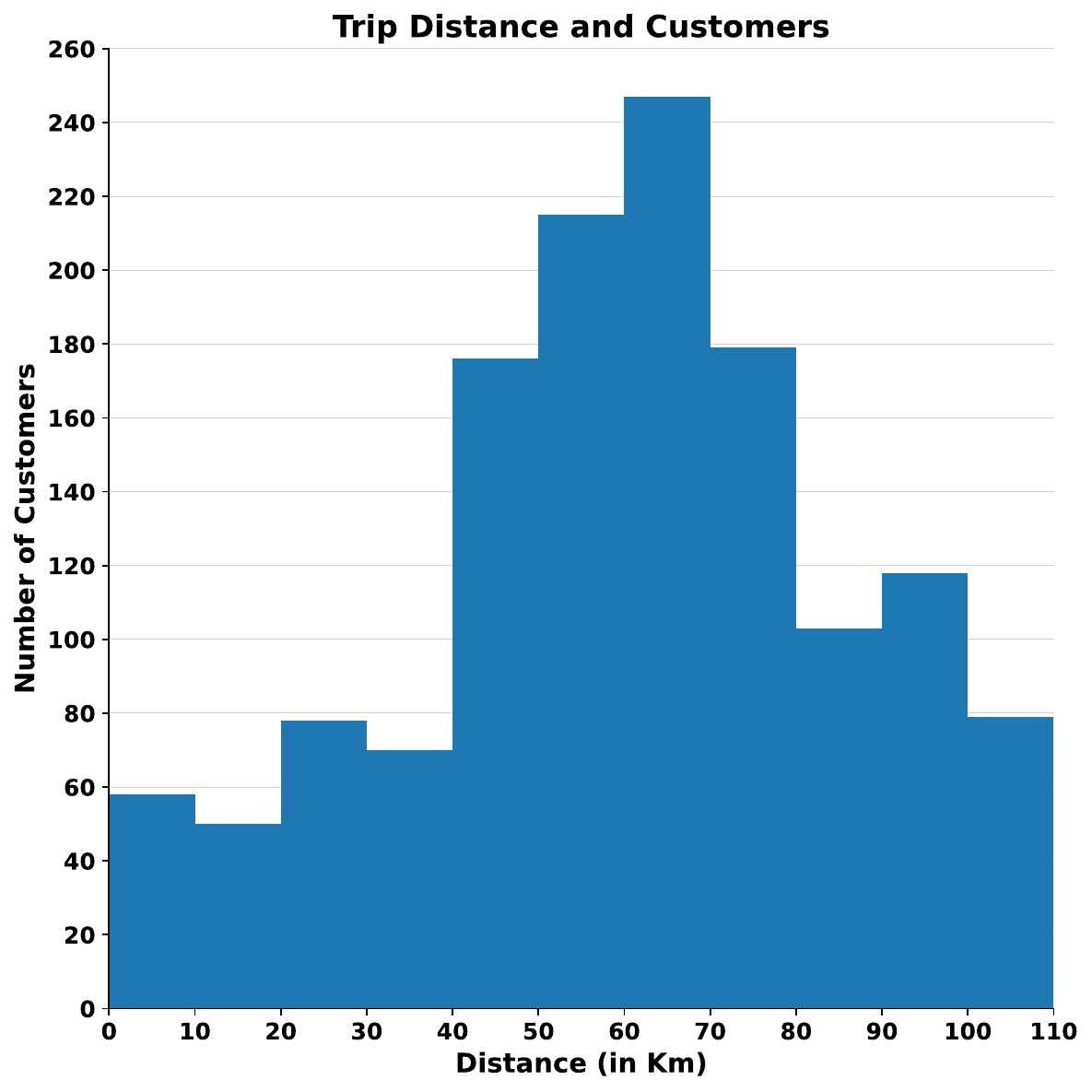} & 
        \includegraphics[width=0.28\textwidth]{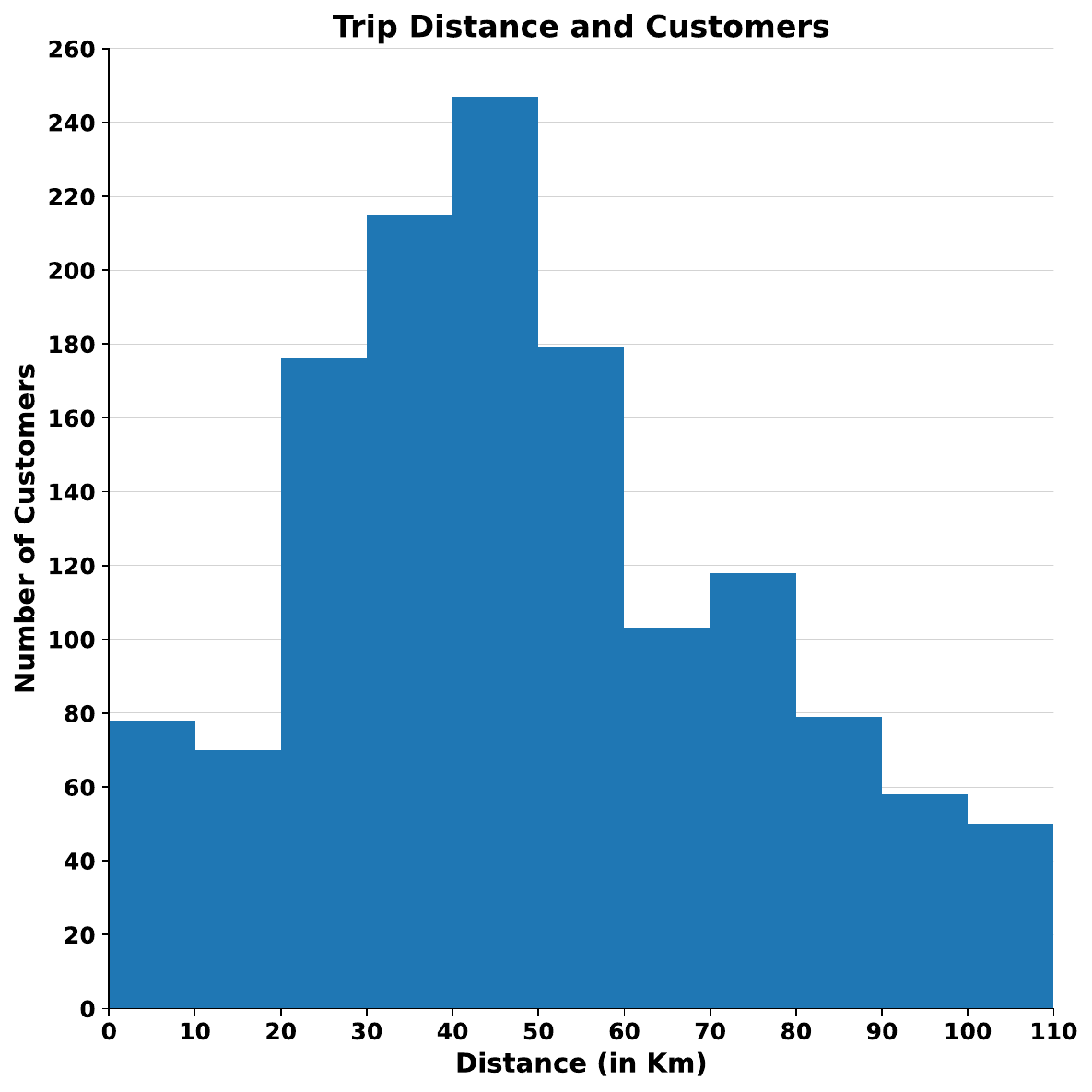} \\
        \midrule
        \textbf{Q:} What was the ratio of girls named “Isla” to girls named “Amelia” in 2012 in the UK? & 
        \textbf{Q:} What was the ratio of girls named “Olivia” to girls named “Amelia” in 2012 in the UK? & 
        \textbf{Q:} What was the ratio of girls named “Isla” to girls named “Olivia” in 2011 in the UK? \\ 
        \textbf{A:} 1 to 1, \colorbox{green!20}{1 to 2}, 1 to 3, 1 to 4 & 
        \textbf{A:} \colorbox{green!20}{1 to 1}, 1 to 2, 1 to 3, 1 to 4 & 
        \textbf{A:} 1 to 1, 1 to 2, \colorbox{green!20}{1 to 3}, 1 to 4 \\ 
        \includegraphics[width=0.28\textwidth]{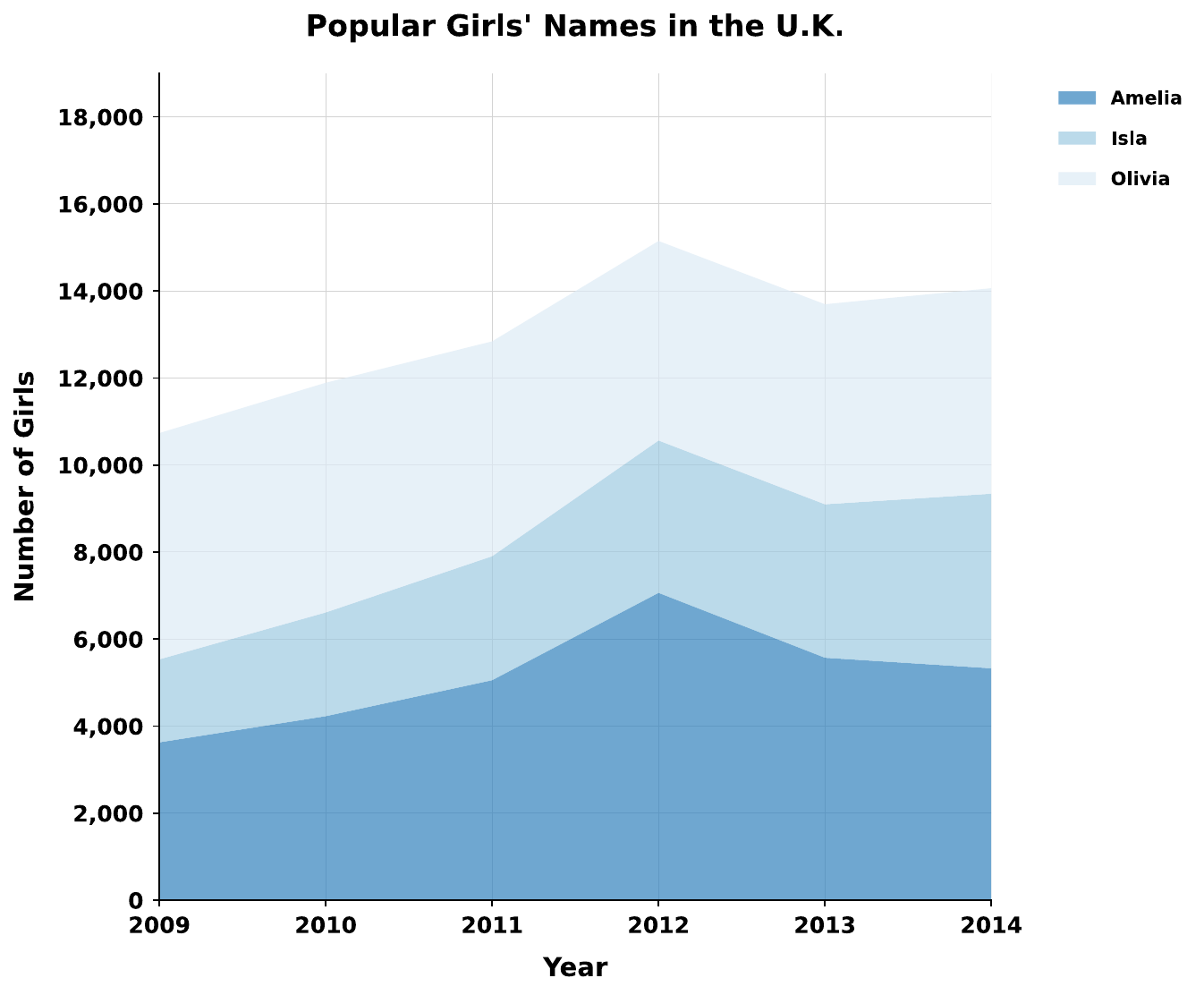} & 
        \includegraphics[width=0.28\textwidth]{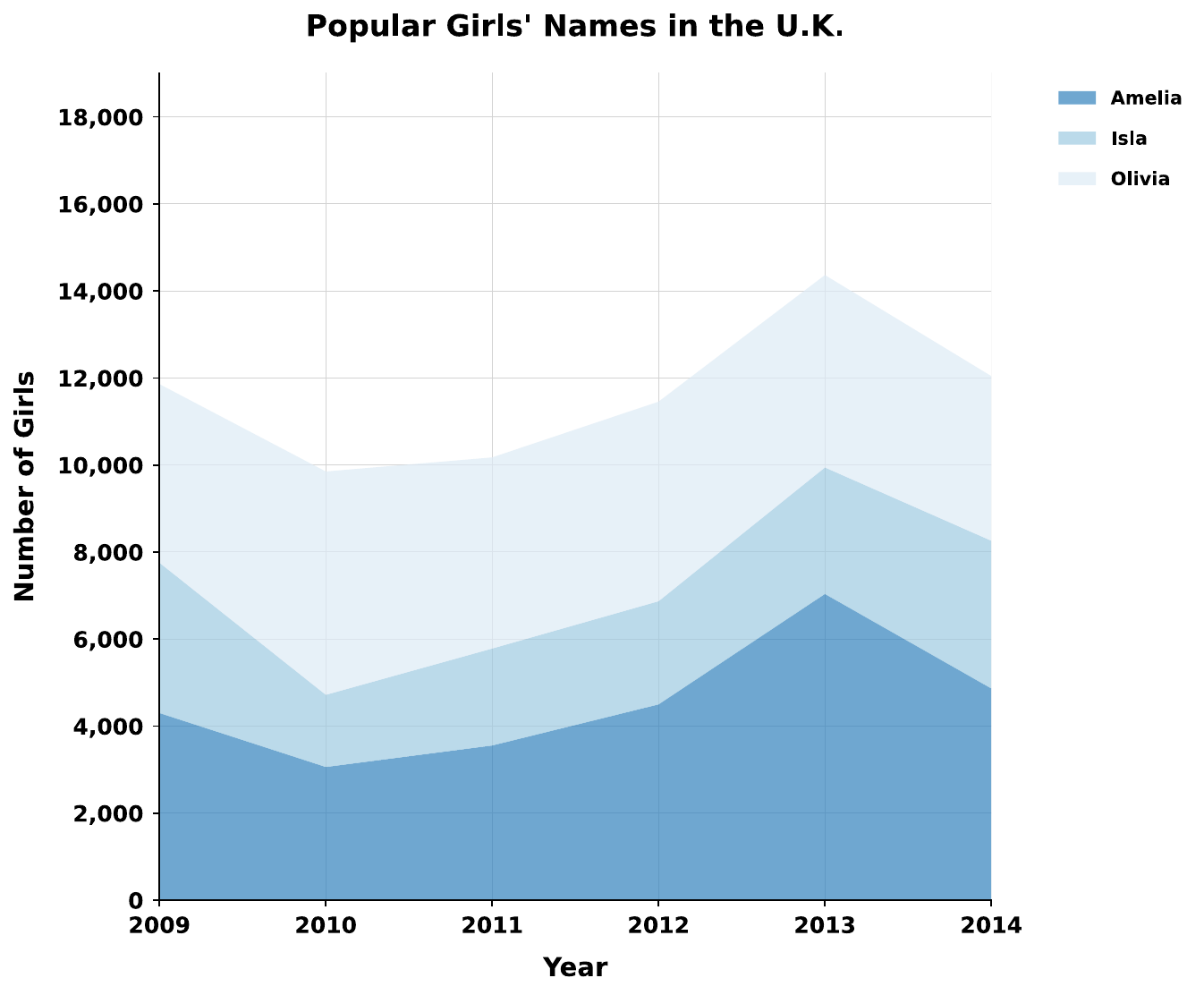} & 
        \includegraphics[width=0.28\textwidth]{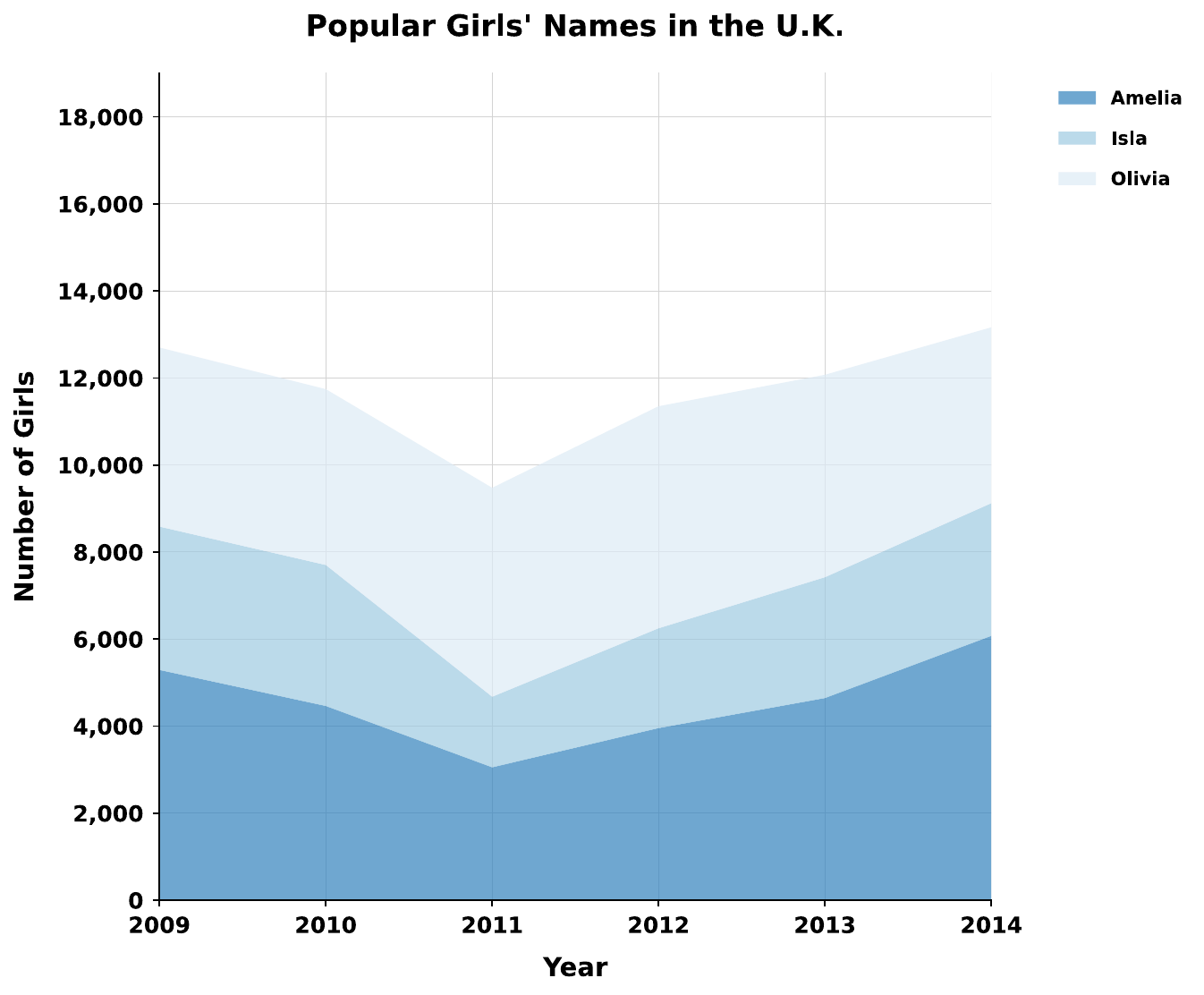} \\
    \end{tabular}
\end{table*}